\documentclass[a4paper,11pt]{article}
\pdfoutput=1 

\usepackage{jcappub} 

\usepackage{graphicx}
\usepackage{amsmath}
\usepackage{amssymb}
\usepackage{rotating} 
\usepackage{xspace}
\usepackage{color}
\usepackage{booktabs}
\usepackage[normalem]{ulem}

\newcommand{\bfx}{{\bf x}}
\newcommand{\bfv}{{\bf v}}
\newcommand{\Msun}{{\rm M}_\odot}

\title{Velocity-dependent J-factors for annihilation radiation from cosmological simulations}

\author[a]{Erin Board,}
\author[b]{Nassim Bozorgnia,}
\author[a]{Louis E. Strigari,}
\author[c]{Robert J. J. Grand,}
\author[d]{Azadeh Fattahi,}
\author[d]{Carlos S. Frenk,}
\author[e]{Federico Marinacci,}
\author[f]{Julio F. Navarro,}
\author[d]{and Kyle A. Oman}

\affiliation[a]{Department of Physics and Astronomy, \\
Mitchell Institute for Fundamental Physics and Astronomy, \\
Texas A$\&$M University, College Station, TX 77843, USA}
\affiliation[b]{York University, Department of Physics and Astronomy,\\
4700 Keele Street, Toronto, Ontario M3J 1P3, Canada}
\affiliation[c]{Max-Planck-Institut f\"{u}r Astrophysik,\\
Karl-Schwarzschild-Str. 1, D-85748, Garching, Germany}
\affiliation[d]{Institute for Computational Cosmology, Durham University,\\
South Road, Durham DH1 3LE, UK}
\affiliation[e]{Department of Physics and Astronomy ``Augusto Righi'',\\ 
University of Bologna, via Gobetti 93/2, 40129 Bologna, Italy}
\affiliation[f]{Department of Physics and Astronomy, University of Victoria,\\ Victoria, BC V8P 5C2, Canada}

\abstract{We determine the dark matter pair-wise relative velocity distribution in a set of Milky Way-like halos in the Auriga and APOSTLE simulations. Focusing on the smooth halo component, the relative velocity distribution is well-described by a Maxwell-Boltzmann distribution over nearly all radii in the halo. We explore the implications for velocity-dependent dark matter annihilation, focusing on four models which scale as different powers of the relative velocity: Sommerfeld, s-wave, p-wave, and d-wave models. We show that the ${\cal J}$-factors scale as the moments of the relative velocity distribution, and that the halo-to-halo scatter is largest for d-wave, and smallest for Sommerfeld models. The ${\cal J}$-factor is strongly correlated with the dark matter density in the halo, and is very weakly correlated with the velocity dispersion. This implies that if the dark matter density in the Milky Way can be robustly determined, one can accurately predict the dark matter annihilation signal, without the need to identify the dark matter velocity distribution in the Galaxy. 
}

\begin{document}
\maketitle
\flushbottom

\section{Introduction}
\label{sec:intro}
Indirect dark matter (DM) searches aim to identify Standard Model (SM) particles that are produced when DM particles annihilate with one another in astronomical environments. Electrons, neutrinos, and photons are stable SM particles that experiments are able to detect. The flux of SM particles from a system depends on the strength of the annihilation cross section, and the phase-space distribution of DM within the system. The astrophysical dependence of the annihilation rate is encapsulated in a quantity typically denoted in the literature as the \emph{${\cal J}$-factor}.   

For DM particles with mass $\sim 10-1000$ GeV, the strongest bounds on the DM annihilation cross section have been obtained through observation of dwarf galaxies by gamma-ray observations such as the Fermi-LAT~\cite{Ahnen:2016qkx,Ackermann:2015zua,Fermi-LAT:2016uux}. Combining the limits from  all  dwarf galaxies  with  high-quality  stellar kinematic data,  these  bounds  reach  the  cosmologically-motivated thermal relic cross section regime over this entire mass range. For higher values of the DM mass, $\gtrsim 1$ TeV, the leading bounds  come  from  observations  of  dwarf galaxies  by  H.E.S.S.~\cite{H.E.S.S.:2020jez} and HAWC~\cite{Albert:2017vtb}. Bounds over this entire mass range may also be obtained from the inner Milky Way (MW) galaxy, though contamination from astrophysical sources make these bounds more difficult to interpret (see e.g Ref.~\cite{Leane:2020liq} for a recent review).  

All these strong bounds on the DM annihilation cross section assume that the cross section is dominated by the velocity independent, s-wave component, and is therefore independent of velocity. If the annihilation cross section is velocity dependent, as in the cases of p-wave, d-wave, or Sommerfeld models, the ${\cal J}$-factor must account for this velocity dependence by incorporating the full dark matter velocity distribution~\cite{MarchRussell:2008tu, MarchRussell:2008yu,Robertson:2009bh,Ferrer:2013cla,Boddy:2017vpe,Zhao:2017dln,Petac:2018gue,Boddy:2018ike,Lacroix:2018qqh,Boddy:2019wfg,FIRE-Jfactor, Hisano:2011dc}. Cross section limits from dwarf spheroidal galaxies have been extended to these velocity-dependent models for the aforementioned annihilation channels~\cite{Zhao:2017dln,Boddy:2019wfg}. These constitute the most stringent limits on  velocity-dependent models. In addition to these bounds from dwarf galaxies, there have been initial explorations of the impact of velocity-dependent dark matter annihilation on the signal from the Galactic center~\cite{Boddy:2017vpe,Johnson:2019hsm}, and from dark matter subhalos~\cite{Boddy:2019qak}.  

The above studies of velocity-dependent DM annihilation rely on simplified analytic models for the DM phase space distribution. While convenient because of their analytic properties, these simplified models need to be tested against the corresponding DM distribution of  MW-like galaxies extracted from  cosmological simulations of galaxy formation. In this paper, we perform the first analysis of the ${\cal J}$-factor in velocity-dependent annihilation cross section models using state-of-the-art simulations of MW-like galaxies. For our study, we use the Auriga magneto-hydrodynamical simulations of galaxy formation~\cite{Grand:2016mgo}, as well as the APOSTLE hydrodynamical simulations~\cite{Sawala2015, 2015arXiv150703643F}. We focus on the expected signal from the MW galaxy, for the first time determining the DM relative velocity distribution from cosmological simulations. From this distribution we determine the velocity-dependent ${\cal J}$-factors for p-wave, d-wave and Sommerfeld annihilation cross section models. 

The paper is organized as follows. In section~\ref{sec:simulations} we discuss the simulations that we use and our criteria for selecting MW-like galaxies. In section~\ref{sec:properties} we determine the density profiles and the relative velocity distributions from our set of simulations. In section~\ref{sec:jfactors} we introduce the formalism for computing the ${\cal J}$-factors for the different DM annihilation models we consider. In section~\ref{sec:results} we present the results for the ${\cal J}$-factors of our selected MW-like galaxies for the smooth halo component. Finally, in section~\ref{sec:conclusions} we discuss our results and summarize our conclusions. In Appendices~\ref{app:fit} and \ref{app:components} we present additional material regarding the DM velocity distributions.

\section{Simulations and halo selection} 
\label{sec:simulations}

In this work we use two different sets of hydrodynamical simulations of MW-mass halos from the Auriga~\citep{Grand:2016mgo} and the APOSTLE~\cite{Sawala:2014xka, 2015arXiv150703643F}
projects, which we discuss in this section.

The Auriga simulations~\citep{Grand:2016mgo} include a suite of thirty magneto-hydrodynamical zoom simulations of isolated MW mass halos, selected from a $100^3$~Mpc$^3$ periodic cube (L100N1504) from the EAGLE project~\cite{Schaye2015,Crain2015}. The simulations were performed using the moving-mesh code Arepo~\citep{Springel2010} and a galaxy formation subgrid model which includes star formation, feedback from supernovae and active galactic nuclei, metal-line cooling, and background UV/X-ray photoionisation radiation~\cite{Grand:2016mgo}. The cosmological parameters used for the simulations are from Planck-2015~\citep{Planck2015} measurements: $\Omega_{m}=0.307$, $\Omega_b=0.048$, $H_0=67.77~{\rm km~s^{-1}~Mpc^{-1}}$. In this work we use the standard resolution level (Level 4) of the simulations with DM particle mass, $m_{\rm DM}=3 \times 10^5~\Msun$, baryonic mass, $m_b=5\times10^4~\Msun$, and  Plummer equivalent gravitational softening of $\epsilon =370$~pc~\citep{Power:2002sw,Jenkins2013}.

The APOSTLE simulations~\cite{Sawala:2014xka, 2015arXiv150703643F} use the same code as the EAGLE project~\citep{Schaye:2015,Crain:2015} with the EAGLE reference model Ref-L100N1504 calibration, applied to zoom simulations of Local Group analogue systems, which contain two MW-mass halos.  The EAGLE
simulations use a modified version of the \textsc{P-gadget3} Tree SPH code~\cite{Springel:2008b}, the {\sc anarchy} version of SPH~\cite{Schaye2015, Schaller:2015b}, and a galaxy formation subgrid model that includes metal-line cooling,
photoionisation, star formation, and feedback from star formation and active galactic nuclei. The cosmological parameters are from WMAP-7: $\Omega_{m}=0.272$, $\Omega_{b}=0.0455$, $h=0.704$. We use twelve APOSTLE volumes simulated at similar
resolution to EAGLE Recal-L025N0752, which we refer to as AP-L2 (i.e.~Level 2 or medium resolution). At this resolution, the DM particle mass, $m_{\rm DM} \simeq 5.9\times10^{5}~\Msun$, the initial gas particle mass, $m_g \simeq 1.3\times10^{5}~\Msun$, and $\epsilon =308$~pc. Notice that the resolution of the halos extracted from the Auriga Level 4 and AP-L2 simulations used in this work are comparable.

All simulated halos have a dark-matter-only (DMO) counterpart which share the same initial conditions as the hydrodynamical runs, but galaxy formation processes are ignored and all the particles are treated as collisionless. In what follows we shall refer to halos in the hydrodynamics simulations as either the Auriga or APOSTLE halos and to those in the DMO simulations as DMO halos.

For the analysis in this work, only DM particles bound to the main halo identified by the SUBFIND algorithm~\cite{Springel:2000qu} are considered. At the end of section~\ref{sec:results}, we briefly discuss how our results change if we include DM particles bound to subhalos.

\subsection{Selection of Milky Way-like galaxies}

Simulated \emph{MW-like} galaxies are usually selected by their virial mass alone. However, to make accurate predictions for the DM distribution throughout the galaxy it is important to apply some additional criteria to select a MW analogue. Here, we specify the criteria we use for selecting MW analogues in the Auriga and APOSTLE simulations. 

The Auriga halos have a virial mass of $M_{\rm 200}=[0.93 - 1.91] \times 10^{12}~\Msun$~\cite{Grand:2016mgo}, which agrees with the observed MW halo mass estimates~(see ref.~\cite{Callingham2019} and references therein). We select the MW analogues by the following criteria introduced in refs.~\cite{Bozorgnia:2016ogo, Bozorgnia:2019mjk}: (i) the stellar mass\footnote{The stellar masses of both the Auriga and APOSTLE halos are calculated from the stars within a spherical radius of 30 kpc from the Galactic center.} of the simulated galaxy falls within the 3$\sigma$ range of the observed  MW stellar mass, $4.5 \times10^{10}<M_{*}/\Msun<8.3 \times10^{10}$~\cite{McMillan:2011wd}, and (ii) the rotation curves of the simulated halos fit well the observed MW rotation curve obtained from ref.~\cite{Iocco:2015xga}. As detailed in ref.~\cite{Bozorgnia:2019mjk}, with these criteria we obtain a total of 10 MW-like Auriga halos. The virial and total stellar masses of these 10 Auriga halos are listed in table \ref{tab:MWlike}. 

 \begin{table}[h!]
    \centering
    \begin{tabular}{|c|c|c|}
      \hline
      Halo Name  & $M_{\rm 200}~[\times 10^{12} \, \Msun]$&  $M_\star~[\times 10^{10} \, \Msun]$ \\
      \hline
      Au2 & 1.91 & 7.65 \\
      Au4 & 1.41 & 7.54 \\
      Au5  & 1.19 & 6.88  \\
      Au7 &  1.12 & 5.27 \\
      Au9 & 1.05 & 6.20 \\
      Au12 & 1.09 & 6.29 \\
      Au19 & 1.21 &  5.72 \\
      Au21 & 1.45 & 8.02\\
      Au22 & 0.93 & 6.10 \\
      Au24  & 1.49 & 7.07\\
      \hline
      AP-V1-1-L2 & 1.64 & 4.88 \\
      AP-V6-1-L2 & 2.15 & 4.48 \\
      AP-S4-1-L2 & 1.47 & 4.23\\
      AP-V4-1-L2 & 1.26 & 3.60\\
      AP-V4-2-L2 & 1.25 & 3.20 \\
      AP-S6-1-L2 & 0.89 & 2.41\\
      \hline
    \end{tabular}
\caption{The virial and stellar masses of the Auriga and APOSTLE MW-like halos, labeled by ``Au-Halo Number'' and ``AP-Volume Number-Halo Number-Resolution Level'', respectively.}
    \label{tab:MWlike}
  \end{table}

The AP-L2 simulations include an initial set of 24 MW-mass halos. Since the stellar masses of the halos in the APOSTLE simulations are slightly smaller than those expected for MW-mass halos~\cite{Schaye:2015}, we slightly relax the criterion on the stellar mass to find the APOSTLE MW-like galaxies. In particular, we select the simulated galaxies with  stellar mass in the range of $2.4 \times10^{10}<M_{*}/\Msun<8.3 \times10^{10}$, and a rotation curve which agrees with the observed MW rotation curve~\cite{Iocco:2015xga}. With these criteria, we obtain a total of 6 MW-like AP-L2  halos. The virial and  stellar masses of these halos are listed in table \ref{tab:MWlike}. 

\section{Properties of MW analogues}
\label{sec:properties}

In this section we discuss the properties of our sample of MW analogues, with a specific focus on the DM density profiles and the relative velocity distributions. Our determination of the DM relative velocity distribution is the first of its kind for MW analogues in cosmological simulations. Our analysis is also the first characterization of the DM velocity distribution at locations inside and outside of the Solar position. All prior studies have focused on the velocity distribution in the solar neighborhood and explored the implications for direct DM detection experiments~\cite{Bozorgnia:2019mjk, Bozorgnia:2016ogo, Bozorgnia:2017brl, Kelso:2016qqj, Sloane:2016kyi}. 

\subsection{Dark matter density profiles}

The predicted DM annihilation signal and the ${\cal J}$-factor are sensitive to the DM density profile, so it is important to understand the behavior of these profiles in our MW analogues. To determine the DM density profiles, we assume the halos to be spherically symmetric. This has been shown to be a good assumption for halos in hydrodynamic simulations~\cite{Schaller:2015mua}, since baryons make the DM distribution more spherical in the central parts compared to the distribution obtained from DMO simulations~\cite{Dubinski:1993df, Abadi:2009ve, Bryan:2012mw, Zhu:2015jwa, Prada:2019tim}. 

The sphericity of the halos can  be directly checked in our simulations. We compute the inertia tensor of the DM particles within four different radii: 2, 8, 20, and 50 kpc from the Galactic center, in Auriga and APOSTLE MW-like halos and their DMO counterparts. The sphericity is defined as $s=c/a$, where $c$ and $a$ are respectively the smallest and largest axes of the ellipsoid obtained from the inertia tensor. For a perfect sphere, $c=a$ and $s=1$. We find that for the Auriga MW-like halos the sphericities at 2, 8, 20, and 50 kpc are in the range of $s (2~{\rm kpc})=[0.66 - 0.89]$, $s (8~{\rm kpc})=[0.72 - 0.86]$, $s(20~{\rm kpc})=[0.71 - 0.88]$, and $s(50~{\rm kpc})=[0.63 - 0.87]$, respectively. As expected, the sphericities are systematically lower for the DMO counterparts, in which $s(2~{\rm kpc})=[0.63 - 0.88]$, $s(8~{\rm kpc})=[0.58 - 0.80]$, $s(20~{\rm kpc})=[0.56 - 0.69]$, and $s(50~{\rm kpc})=[0.49 - 0.70]$. For the APOSTLE MW-like halos, we find $s(2~{\rm kpc})=[0.80 - 0.90]$, $s(8~{\rm kpc})=[0.69 - 0.88]$, $s(20~{\rm kpc})=[0.73 - 0.85]$, and $s(50~{\rm kpc})=[0.71- 0.91]$, while for their DMO counterparts, $s(2~{\rm kpc})=[0.75 -  0.79]$, $s(8~{\rm kpc})=[0.60 - 0.75]$, $s(20~{\rm kpc})=[0.54 - 0.75]$, and $s(50~{\rm kpc})=[0.53 - 0.78]$.

We extract the spherically-averaged DM density profiles from the mass enclosed in consecutive spherical shells of different widths from the Galactic center, containing 2,000 DM particles within each shell. Our choice of 2,000 DM particles per shell optimizes the calculation time of the ${\cal J}$-factors discussed in section~\ref{sec:jfactors}. In order to calculate accurately the DM density profile, it is important to  choose the location of the halo center carefully. We determine the center of each halo using the shrinking sphere method~\cite{Power:2002sw}. This is an iterative technique in which we start by calculating the center of mass of the DM particles within the virial radius, and then recursively shrink the radius of the sphere. At each step of the iteration the center of the halo is reset to the last computed barycenter and the radius of the sphere is reduced by 5\%. This process continues until 1000 DM particles are contained within the sphere.

A second issue which is important in determining the DM density profile is the resolution limit. The thorough resolution study of Ref.~\cite{Power:2002sw} suggests a convergence radius at which the
integrated mass is converged within $\sim 10\%$, i.e.~the so-called \emph{Power radius}, $R_{\rm P03}$, based on the two-body relaxation timescale of the DM particles. The criterion can be written as:
\begin{equation}
0.6 \leq \frac{\sqrt{200}}{8} \sqrt{\frac{4\pi\rho_{\rm crit}}{3m_{\rm DM}}} \frac{\sqrt{N}}{\ln{N}} R_{\rm P03}^{3/2},
\label{eq:power_radius}
\end{equation}
where N is the number of particles with mass $m_{\rm DM}$ enclosed within $R_{\rm P03}$, and $\rho_{\rm crit} = 3H^2/8\pi G$ is the critical density~\cite{Schaller_2015}. For the cosmological parameters used in the simulations, we have $\rho_{\rm crit}(z=0) = 127.49$ ~$\Msun$~kpc$^{-3}$ and $137.58$ $\Msun$~kpc$^{-3}$ for Auriga and APOSTLE simulations, respectively. Solving eq.~\eqref{eq:power_radius} for each of the halos in the DMO simulations, we find the Power radius to be in the range of $R_{\rm P03} = [1.14 - 1.29]$ kpc and $R_{\rm P03} = [1.41 - 1.59]$ kpc for the Auriga and APOSTLE DMO simulations, respectively. The concept of numerical convergence is less clear in simulations containing baryons. For halos in the hydrodynamic simulations, we calculate the Power radius using only the DM particles and multiplying their mass by a factor of $\Omega_{m}/\Omega_{\rm DM}$, which corresponds to a halo entirely made of DM particles. We find that the Power radius is in the range of $R_{\rm P03} = [0.94 - 1.07]$ kpc and $R_{\rm P03} = [1.33 - 1.45]$ kpc for Auriga and APOSTLE MW-like halos, respectively. The average Power radius is $R_{\rm P03} = 0.98$ kpc and $1.41$ kpc for the 10 Auriga and 6 APOSTLE MW-like halos, respectively.

\begin{figure}[h]
    \centering
    \includegraphics[width=0.495\textwidth]{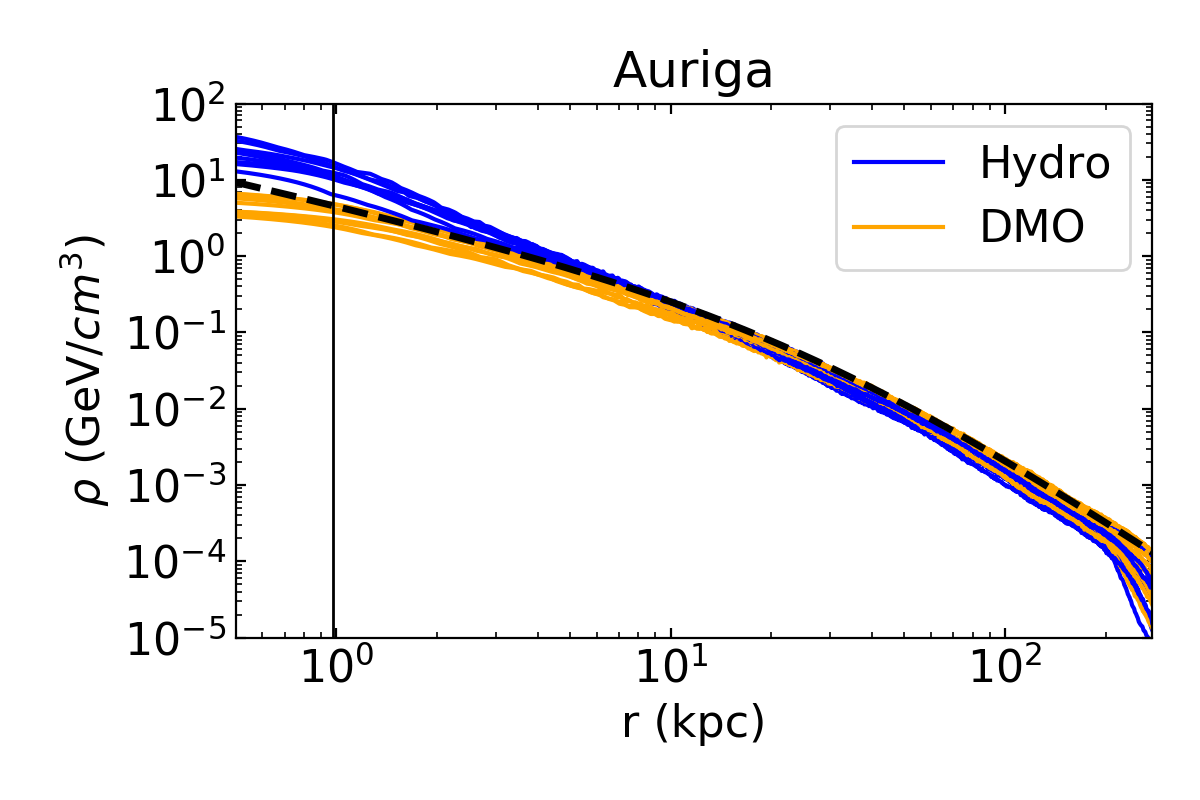}
    \includegraphics[width=0.495\textwidth]{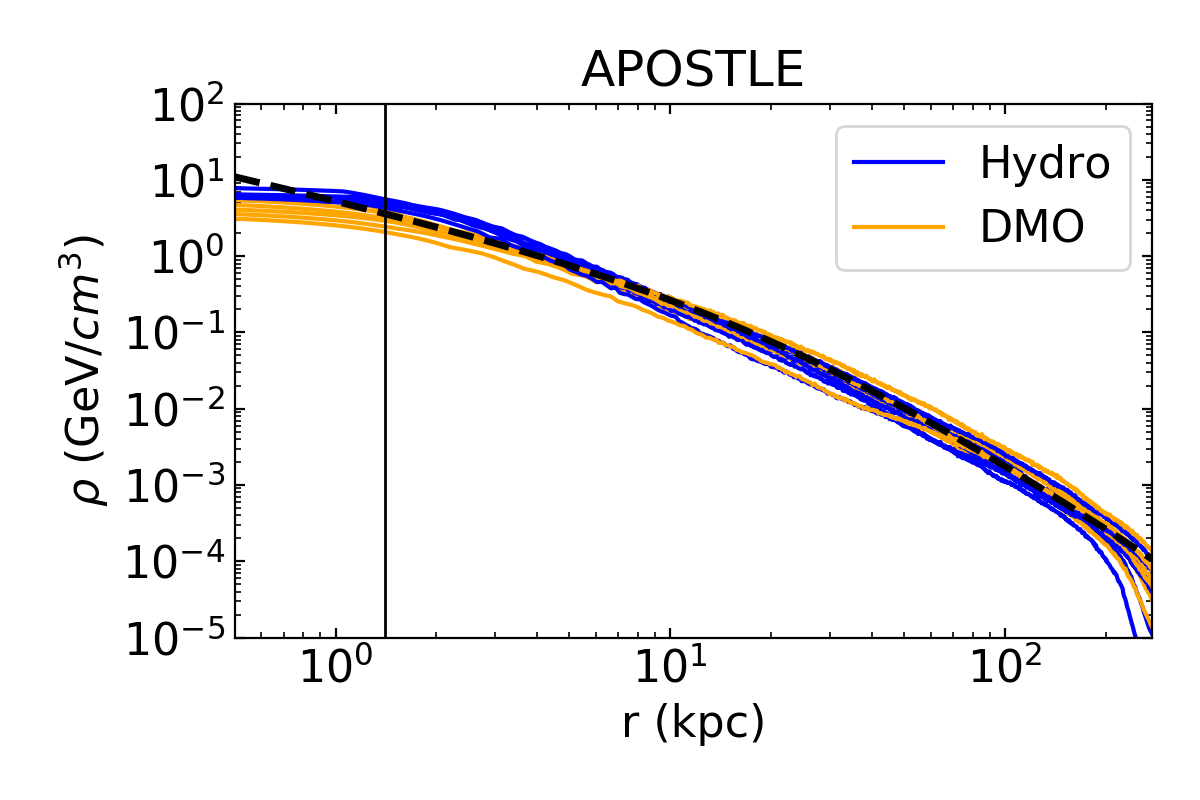}
    \caption{DM Density profiles for the Auriga (left panel) and APOSTLE (right panel) MW-like halos (blue) and their DMO counterparts (yellow). The dashed black curves specify the best fit NFW profile for Auriga halo Au2 in the left panel and APOSTLE halo AP-V4-1-L2 in the right panel. The vertical lines mark the average Power radius for the Auriga and APOSTLE MW-like halos in the left and right panels, respectively.}
    \label{fig:AurigaDensityProfiles}
\end{figure}

Using the methodology described above, figure~\ref{fig:AurigaDensityProfiles} shows the DM density profiles for our MW analogues in the Auriga (left panel) and APOSTLE (right panel) simulations. As expected, at large radii, there is essentially complete agreement between the DM density profiles of the DMO and the hydrodynamic simulations. At small radii, inside the expected location of the Solar circle, the trend is for the halos in the hydrodynamic simulations to have steeper profiles compared to the DMO. This is a result of the contraction of the DM halo as a response to the presence of baryons in the inner parts of the halo~\cite{Cautun:2019eaf, Callingham:2020ips}. The steepening of the hydrodynamic profiles compared to their DMO counterparts is more pronounced for the Auriga halos compared to the APOSTLE halos. This is due to the smaller stellar masses of the APOSTLE halos, which leads to less contraction of the halos in APOSTLE compared to Auriga. For comparison, the best fit Navarro–Frenk–White (NFW) profile for the Auriga halo Au2 in the left panel and APOSTLE halo AP-V4-1-L2 in the right panel are shown as dashed black curves in figure~\ref{fig:AurigaDensityProfiles}.

\subsection{Relative velocity distributions}
\label{subsec:relative velocity distributions}

We now determine the DM pair-wise velocity distributions, to which we refer in what follows as the DM relative velocity distributions. We begin by establishing our notation. Define $f(\bfx, \bfv)$ such that $f(\bfx, \bfv) ~d^3 \bfx~ d^3 \bfv$ is the mass of DM particles within a phase space volume $\bfx + d^3 \bfx$ and $\bfv + d^3 \bfv$. The position vector $\bfx$ and the velocity vector $\bfv$ are defined in the rest frame of the galaxy. In these expressions and those below, bold-face quantities represent vectors with components given by the three spatial and velocity components of a DM particle. At a position $\bfx$ in the halo, we write the probability distribution of DM velocities as 
\begin{equation} 
P_{\bfx} (\bfv) = \frac{f(\bfx, \bfv)}{\rho(\bfx)},
\end{equation}
where the DM density at $\bfx$ is normalized as
\begin{equation} 
\rho ({\bfx}) = \int f ({\bfx},{\bfv}) d^3 {\bfv}. 
\label{eq:df} 
\end{equation} 

At a position $\bfx$, we are interested in the probability that a DM particle 1 has velocity $\bfv_1$ in the range $\bfv_1 + d^3 \bfv_1$ times the probability that a DM particle 2 has velocity $\bfv_2$ in the range $\bfv_2 + d^3 \bfv_2$,
\begin{equation}
P_\bfx (\bfv_1) d^3 {\bfv_1} 
P_\bfx (\bfv_2) d^3 {\bfv_2}. 
\end{equation}
The individual particle velocities may be written in terms of the center-of-mass velocity, $\bfv_{\rm cm}$, and the relative velocity, $ \bfv_{\rm rel} \equiv \bfv_2 - \bfv_1$, as $\bfv_1 = \bfv_{\rm cm} + \bfv_{\rm rel}/2$ and $\bfv_2 = \bfv_{\rm cm} - \bfv_{\rm rel}/2$. Using the fact that the magnitude of the jacobian of the transformation $d^3 {\bfv_1} d^3 {\bfv_2} \rightarrow d^3 {\bfv_{\rm cm}} d^3 {\bfv_{rel}}$ is unity, and integrating over ${\bfv_{\rm cm}}$, we then obtain a general expression for the distribution of relative velocities at a position \bfx,  
\begin{equation} 
P_{\bfx}(\bfv_{\rm rel}) = \int P_\bfx (\bfv_1 = \bfv_{\rm cm} + \bfv_{\rm rel}/2) P_\bfx (\bfv_2 = \bfv_{\rm cm} - \bfv_{\rm rel}/2) ~d^3 \bfv_{\rm cm}.  
\label{eq:fvrel}
\end{equation}

To calibrate our expectations, it is useful to review the prediction for the relative velocity distribution in the case of a pure Maxwellian halo. For Maxwellian halos, at any point in the halo, the DM velocity distribution, $f$, is Gaussian in all three velocity components, with a dispersion in each direction given by $\sigma$. The distribution of velocities is then given by the Standard Halo Model (SHM)~\cite{Drukier:1986tm}, which is the simplest and most commonly adopted model to describe the DM halo. In the SHM, the DM halo is assumed to be spherical and isothermal, and this leads to an isotropic Maxwell-Boltzmann velocity distribution with a most probable speed of $\sqrt{2} \sigma$. In this case, the relative velocity distribution, $P_{\bfx}(\bfv_{\rm rel})$, is also a Maxwellian distribution, but with a one dimensional relative velocity dispersion of $\sqrt{2} \sigma$~\cite{Ferrer:2013cla}.  

The velocity vectors of the simulation particles are determined with respect to the center of each halo. In each spherical shell, we resolve the velocity vectors into three components then subtract the components of the velocities in this basis, being careful to avoid double counting. We then take the modulus of the components of the pairwise relative velocities, which provides an estimate of $P_{\bfx}(\bfv_{\rm rel})$ in each radial shell.

Notice that the relative velocity modulus distribution, $P_{\bfx}(|\bfv_{\rm rel}|)$, is related to the relative velocity distribution, $P_{\bfx}(\bfv_{\rm rel})$, by
\begin{equation} 
P_{\bfx}(|\bfv_{\rm rel}|)= v_{\rm rel}^2 \int P_{\bfx}(\bfv_{\rm rel}) ~d \Omega_{\bfv_{\rm rel}}, 
\end{equation} 
where $d \Omega_{\bfv_{\rm rel}}$ is an infinitesimal solid angle along the direction $\bfv_{\rm rel}$. In each radial shell, $P_{\bfx}(|\bfv_{\rm rel}|)$ is normalized to unity, such that 
\begin{equation} 
\int P_{\bfx}(|\bfv_{\rm rel}|) ~dv_{\rm rel}= 1 
\end{equation} 
and therefore we have $\int P_{\bfx}(\bfv_{\rm rel}) ~d^3 \bfv_{\rm rel}= 1$.

\begin{figure}[t]
    \includegraphics[width=0.495\textwidth]{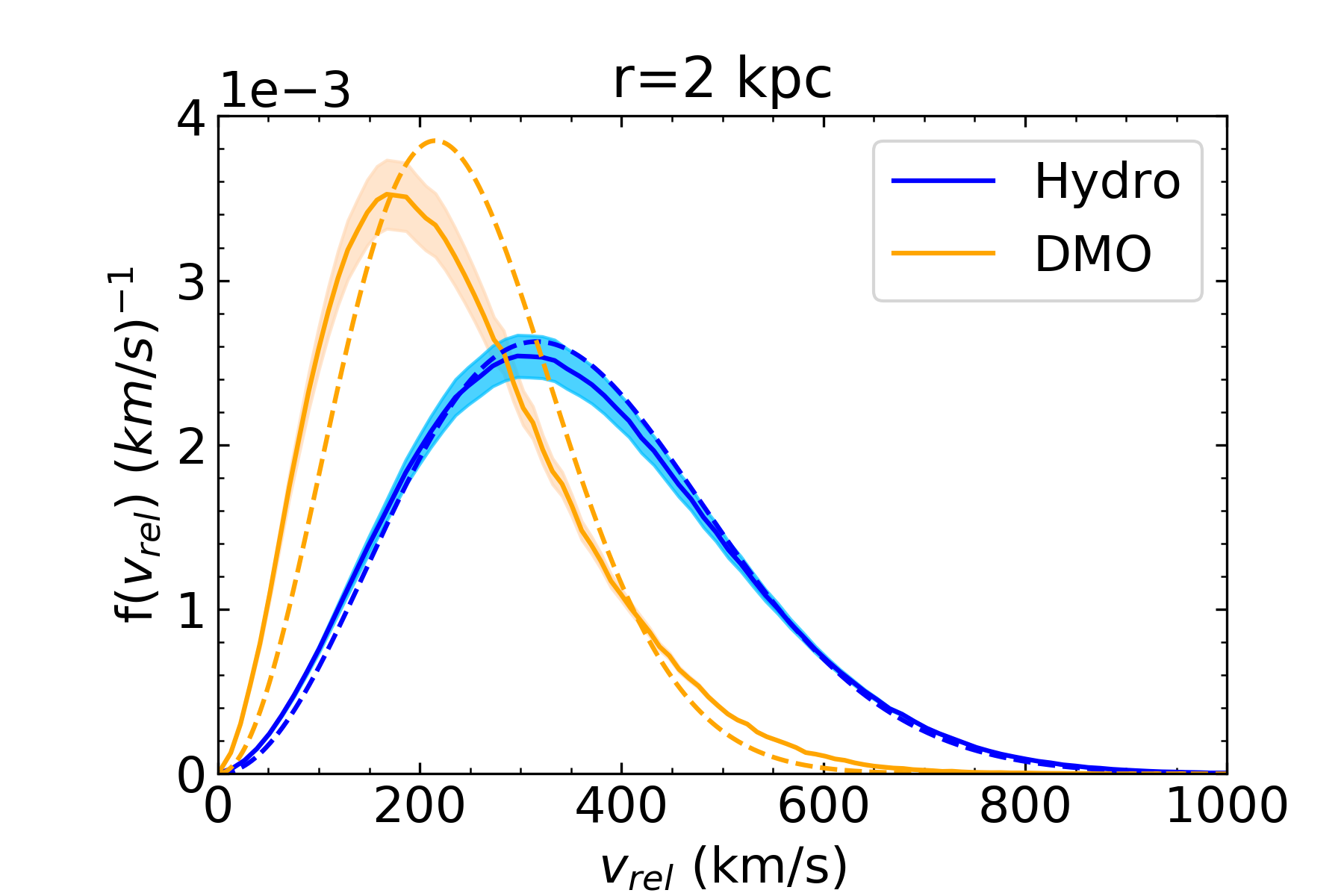}
    \includegraphics[width=0.495\textwidth]{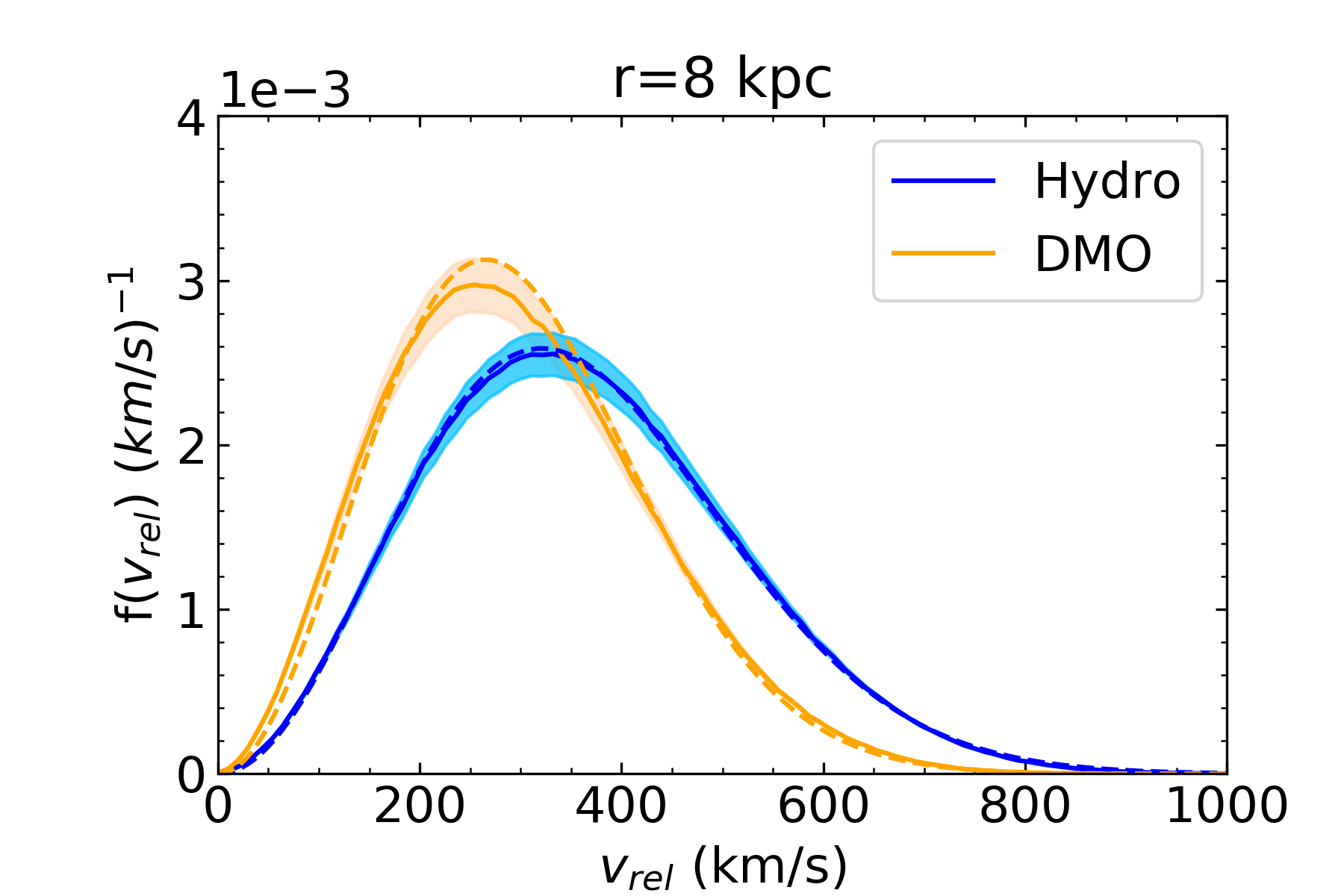}
    \includegraphics[width=0.495\textwidth]{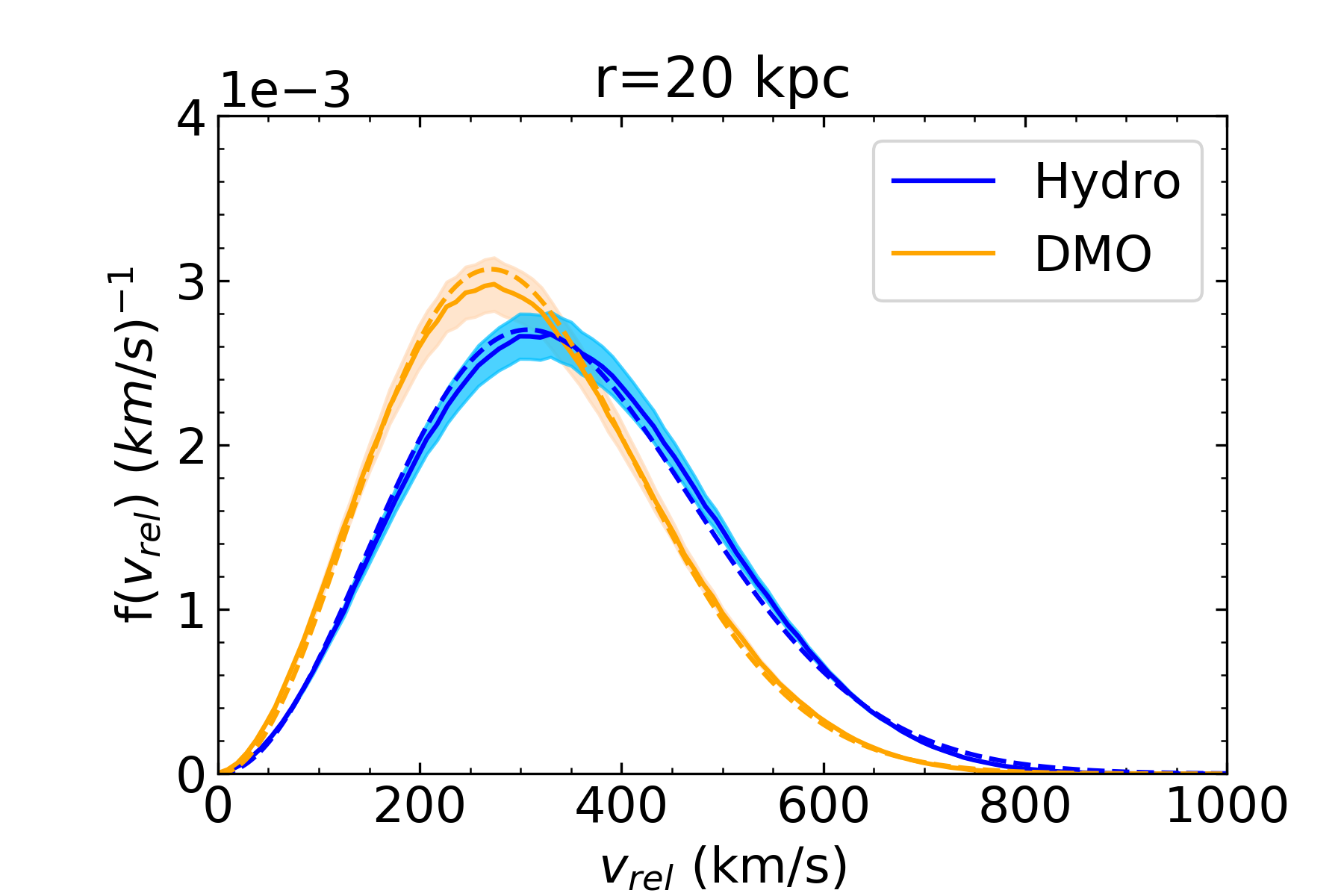}
    \includegraphics[width=0.495\textwidth]{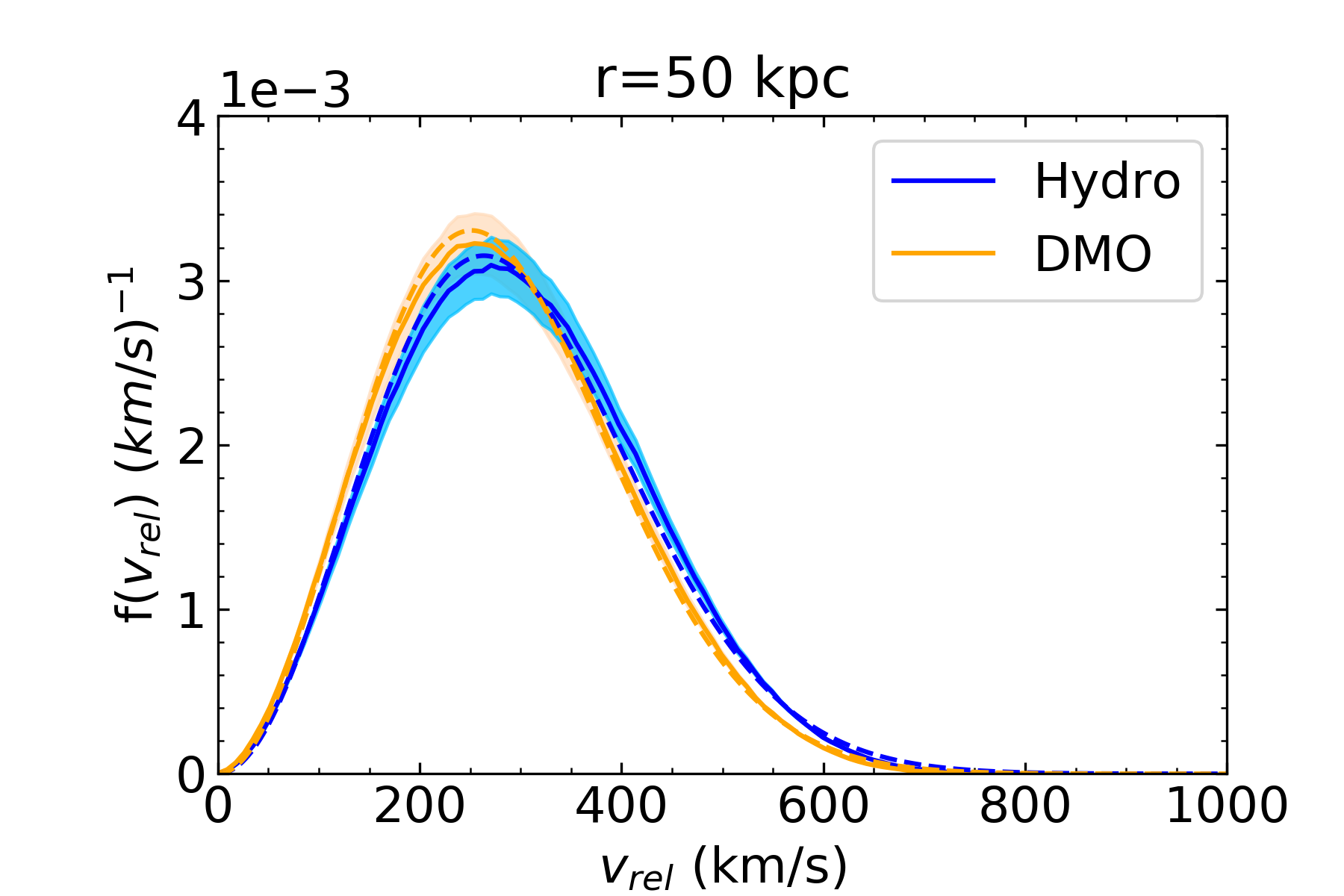}
    \caption{Modulus of the DM relative velocity distributions in the Galactic rest frame for an example  Auriga MW-like halo (blue) and its DMO counterpart (yellow). Each panel shows the distributions at a different Galactocentric radius. The solid curves specify the mean relative speed distributions, while the shaded bands specify the 1$\sigma$ Poisson errors. The  dashed curves represent the corresponding best fit Maxwell-Boltzmann distribution.}
    \label{fig:vreldistributions}
\end{figure}

In figure~\ref{fig:vreldistributions} we show the DM relative velocity modulus distribution in the Galactic rest frame for an example MW-like  Auriga halo and its respective DMO counterpart. For both halos, we show the speed distributions in radial shells near the Galactic center, near the Solar circle, and at two radii well beyond the Solar circle (i.e.~20 and 50 kpc from the Galactic center). The solid blue (orange) curves show the mean speed distribution for the Auriga (DMO) halo, while the shaded bands specify the $1 \sigma$ Poisson error in the speed distributions. 

\begin{figure}[h]
    \includegraphics[width=0.495\textwidth]{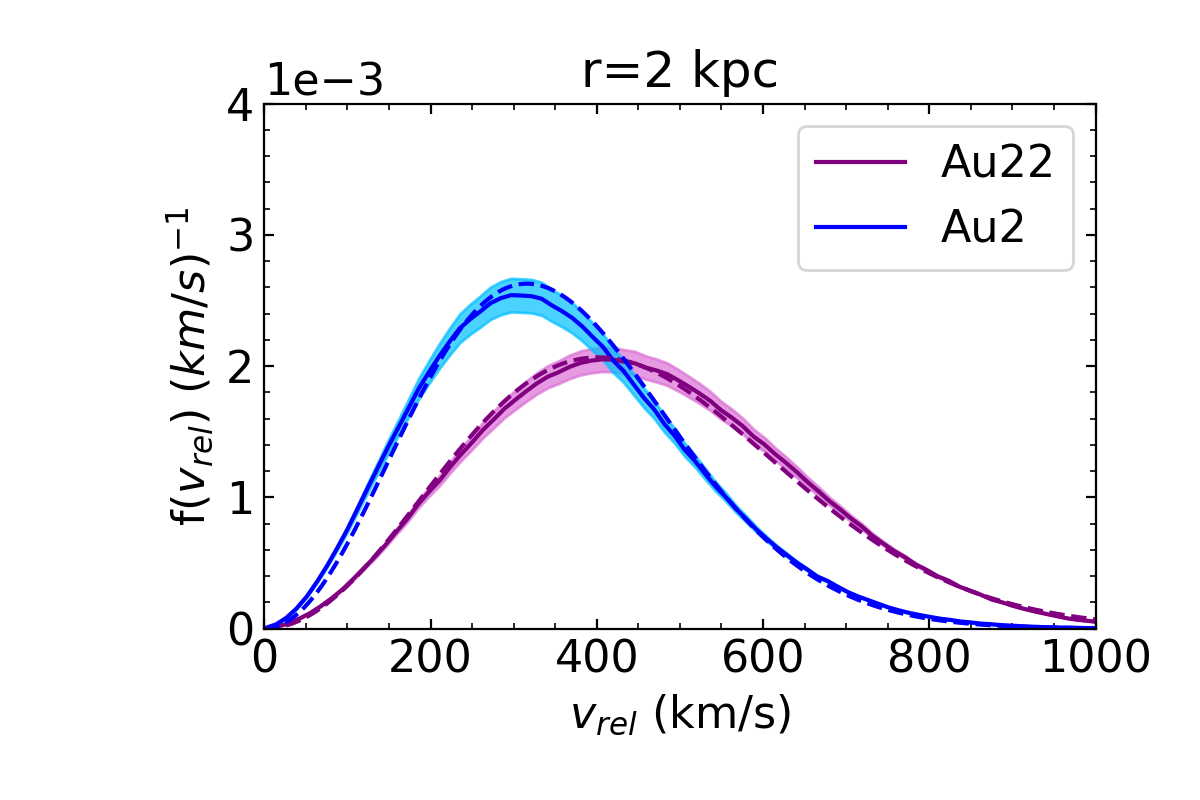}
    \includegraphics[width=0.495\textwidth]{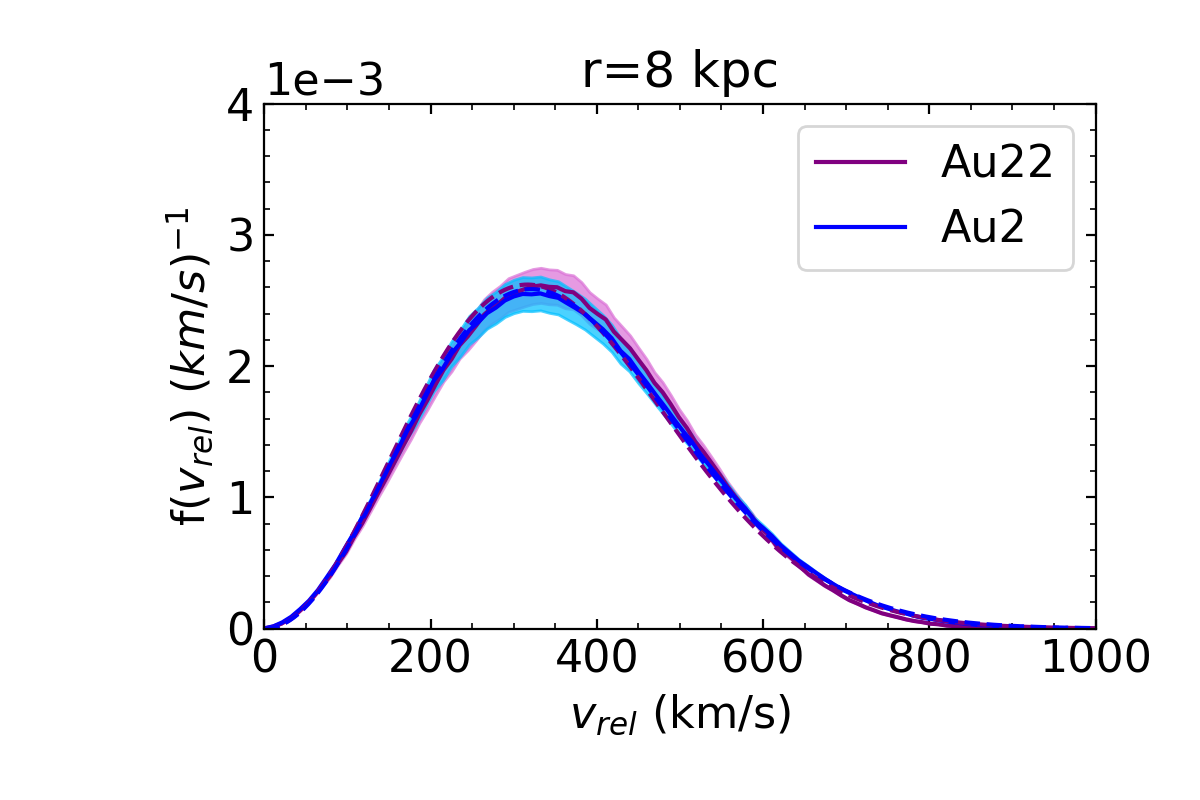}
    \includegraphics[width=0.495\textwidth]{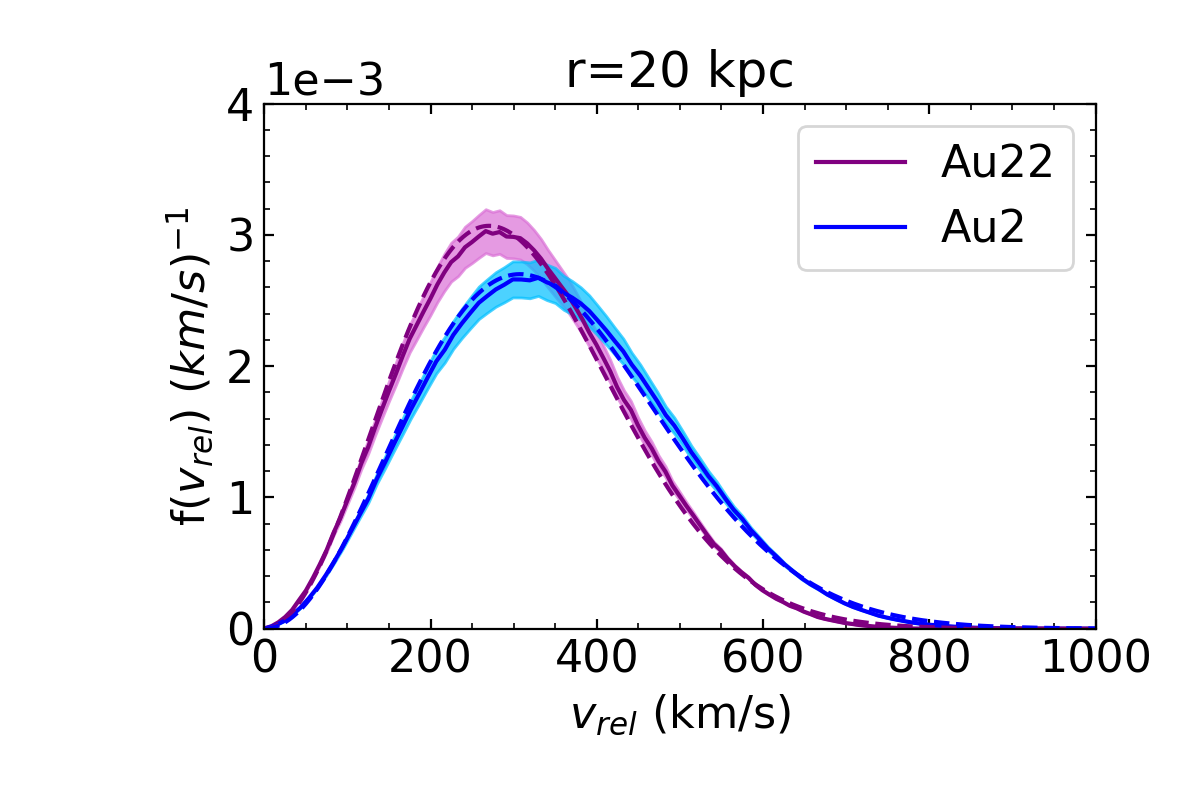}
    \includegraphics[width=0.495\textwidth]{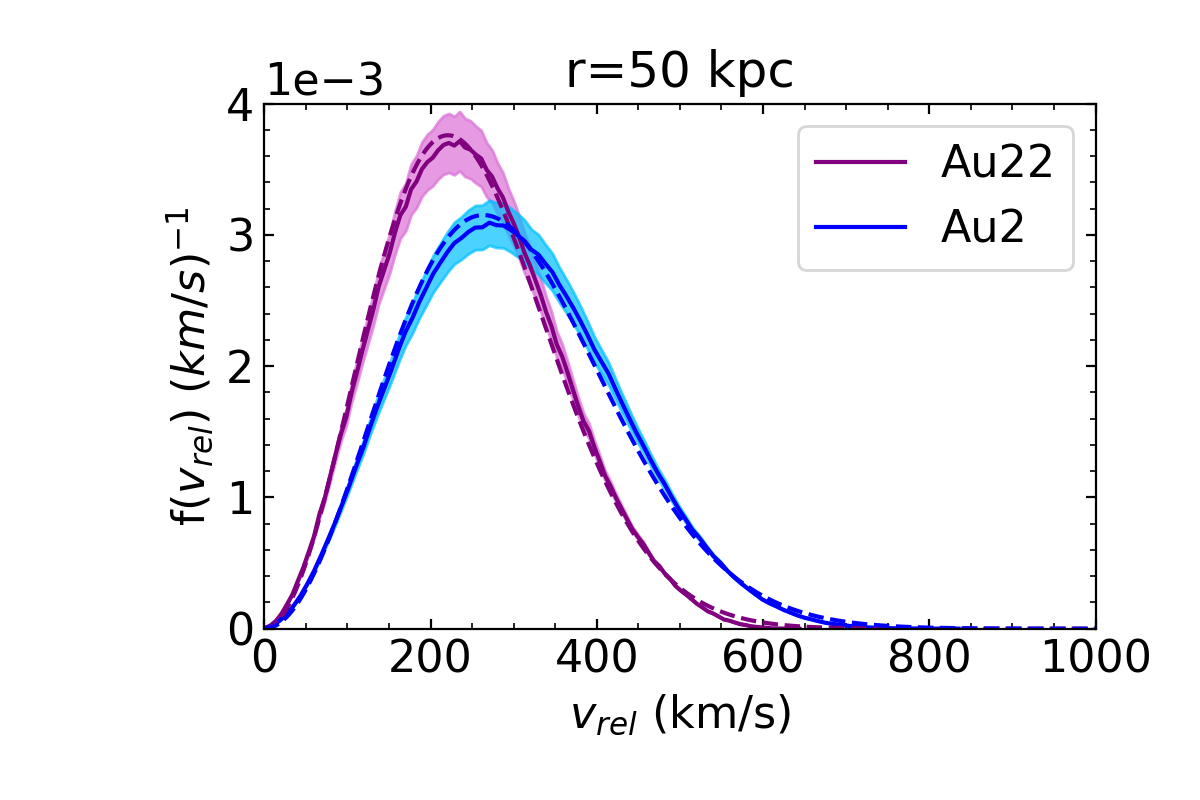}
    \caption{Modulus of the DM relative velocity distribution for the two   Auriga MW-like halos that have the smallest (Au2, blue) and largest (Au22, magenta) peak speeds at 2 kpc. The modulus velocity distributions for the two halos are shown at the same radii as in figure~\ref{fig:vreldistributions}.
    \label{fig:hydrovrel}}
\end{figure}

The method used to define the spherical shells for calculating the density profiles produces varying radial boundaries from halo to halo. In order to effectively compare the relative velocity distributions of different halos at the same radius, we redefine the spherical shells to have fixed radial width progressing outward from the Galactic center. Each spherical shell has radial width of 0.1 kpc, with the number of particles in each shell in the range of $[486-3304]$. The spherical shells of fixed radial width are only used in the calculations shown in figures~\ref{fig:vreldistributions} and \ref{fig:hydrovrel} (also see figure~\ref{fig:hydro vrel components}).

As we can see from figure~\ref{fig:vreldistributions}, including baryons in the simulations results in an increase of the DM relative speed distributions at all radii. This increase is more pronounced in the inner galaxy, and is due to the deepening of the galaxy's gravitational potential when baryons are included in the simulations. This result is consistent with the  local DM speed distributions of MW-like galaxies extracted from other hydrodynamic simulations~\cite{Bozorgnia:2019mjk, Bozorgnia:2017brl, Bozorgnia:2016ogo, Kelso:2016qqj, Sloane:2016kyi}.

Next, we compare the DM relative speed distributions at each radii with a Maxwellian distribution (dashed colored curves in figure~\ref{fig:vreldistributions}). For each halo in the hydrodynamic and DMO simulations, we find the best fit Maxwellian speed distribution, $f(v) \propto v^2 \exp(-v^2/v_0^2)$, where $v_0$ is the best fit peak speed. For the halos in the hydrodynamic simulations, the relative speed distributions are very close to the Maxwellian model at all radii, with an agreement becoming increasingly better as we move further away from the Galactic center. For the DMO halos, the agreement with the Maxwellian model is not as good as is for the hydrodynamic case, though again the agreement gets better at radii further away from the Galactic center. Deviations from the Maxwellian distribution for the DMO halos at small radii are not surprising, since the DM density profiles deviate from the isothermal $r^{-2}$ profile in the central regions of the DMO halos~\cite{Kazantzidis:2003im}. Additionally, the velocity anisotropy of the DMO halos at all radii leads to further deviations from the isotropic Maxwellian distribution. 

In all cases, the DM relative speed distribution at small radii is shifted to smaller relative speeds as compared to the Maxwellian distributions, while at large radii there is a shift to larger relative speeds compared to the Maxwellian. We explore the origins of the shapes of these distributions in the following section. To understand how good the fit is to the Maxwell-Boltzmann distribution, in Appendix~\ref{app:fit} we present the $\chi^2$/dof for all halos at several different radii. 

\begin{figure}[t]
\begin{center}
    \includegraphics[width=0.6\textwidth]{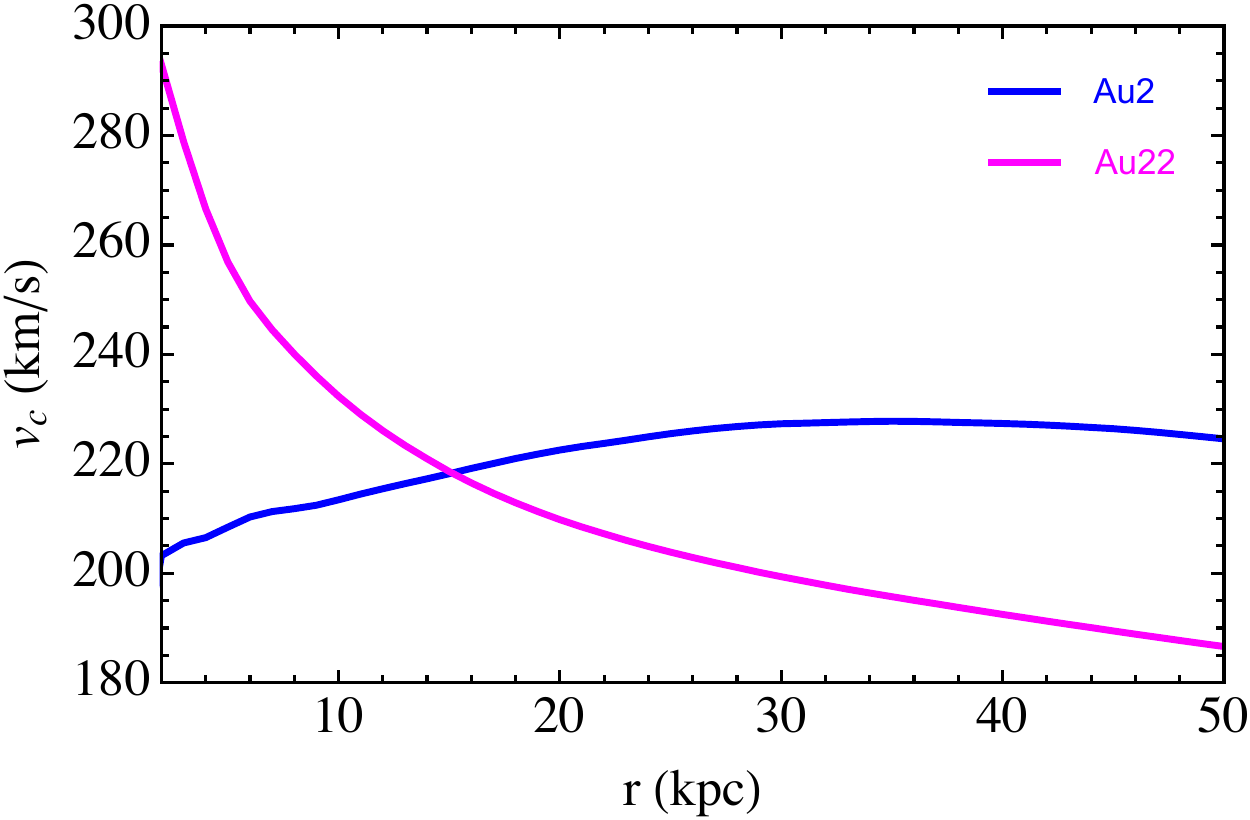}
    \caption{Circular velocity of the two Auriga halos Au2 (blue) and Au22 (magenta) as function of Galactocentric radius.}
    \label{fig:vc}
\end{center}
\end{figure}

To explore the halo-to-halo variation in the DM relative speed distributions of the Auriga MW-like halos, we first examine their rotation curves. The circular velocities for two example Auriga  halos (Au2 and Au22) are shown in figure~\ref{fig:vc}. The total circular velocity of each halo is $v_c(r)=\sqrt{G M(<r)/r}$, where $M(<r)$ is the total mass (DM, stars, and gas) enclosed in a sphere of Galactocentric radius $r$. In figure~\ref{fig:hydrovrel}, we show the relative velocity modulus distributions for the same two halos. These halos have the smallest and largest peak speeds in the radial shell centered at 2 kpc. The four panels show the relative speed distributions of the two halos at different Galactocentric radii. As we move from 2 kpc to 50 kpc from the Galactic center, the relative speed distributions of Au22 is strongly shifted to smaller speeds, while that of Au2 does not show a significant change. This behavior can be understood from the rotation curves of the two halos, shown in figure~\ref{fig:vc}. The circular velocity of Au2 changes slightly with Galactocentric distance, while that of Au22 decreases significantly as we move from 2 kpc to larger radii.

Notice that to extract the relative DM velocity distributions, we calculate the average distribution in each radial shell. We have verified the spherically average velocity distributions we obtained are consistent with those obtained by splitting each radial shell into 8 sections divided evenly about the azimuthal direction of the halo's principal axes. We have also checked our results against a more local method for computing the relative DM velocity distributions, using only the nearest neighbors of each particle. Choosing reasonable aperture sizes to find the neighbors of each particle in each radial shell, we find that the relative velocity distributions and ${\cal J}$-factors are not significantly affected. The difference in all the results of this paper when using this local nearest neighbors method compared to using all particle pairs is at the order of $\sim 10\%$.

\section{J factors} 
\label{sec:jfactors}

Having determined the DM density profiles and the relative velocity distributions for the MW-like halos, we are now in position to determine the velocity-dependent ${\cal J}$-factors. In this section, we lay out the formalism for calculating the ${\cal J}$-factors for each of the annihilation cross section models that we consider. In the formulae presented below, our notation closely follows that of Ref.~\cite{Ferrer:2013cla}. 

\subsection{Annihilation rate}
We begin by defining $\sigma_A$, the DM annihilation cross section to any set of Standard Model particles. The number density of DM particles at position \bfx~is $\rho(\bfx)/m$, where $m$ is the DM particle mass. The flux of DM particles is given by the product of the number density and the modulus of the relative velocity, $v_{\rm rel} \equiv |{\bfv}_{\rm rel}| = | {\bfv}_1 - {\bfv}_2 |$. Multiplying the flux by the DM annihilation cross section and the number density of target DM particles, we obtain the annihilation rate in a volume element $d V$ at the position $\bfx$ in the halo as
\begin{equation} 
\frac{d \Gamma}{dV} = \left[\frac{\rho(\bfx)}{m}\right]^2 \int {\rm d}^3 \bfv_{\rm rel} P_{\bfx} (\bfv_{\rm rel}) (\sigma_A {v}_{\rm rel}). 
\label{eq:dgammadv}
\end{equation} 
We note that the standard definition of the annihilation cross section averaged over the relative velocity distribution is then, 
\begin{equation} 
\langle \sigma_A v_{\rm rel} \rangle(\bfx) =\int {\rm d}^3 \bfv_{\rm rel} P_{\bfx} (\bfv_{\rm rel}) (\sigma_A {v}_{\rm rel}),
\label{eq:sigmavrel} 
\end{equation} 
which in general depends on spatial location $\bfx$. 

\par To determine the annihilation rate, as above we take the DM halo as spherically symmetric. We define a solid angle centered on the Galactic center, $r$ as the distance from the Galactic center to a point in the halo, $R_0$ as the distance from the Sun to the Galactic center, $\ell$ as the distance from the Sun to a point in the halo (i.e.~line of sight), and $\Psi$ as the opening angle between the line of sight $\ell$ and the direction towards the Galactic center. The radial distance from the Galactic center to a point in the halo can then be expressed as $r^2\left(l,\Psi\right)=l^2+R_0^2-2lR_0\cos \Psi$. The annihilation rate along the line of sight is then proportional to 
\begin{equation} 
{\cal J}_s(\Psi) = \int d \ell \, \frac{\langle \sigma_A v_{\rm rel} \rangle}{(\sigma_A v_{\rm rel})_0}  \left[\rho (r(\ell, \Psi))\right]^2. 
\label{eq:Jfactor}
\end{equation} 
which, following Ref.~\cite{Boddy:2019wfg}, we define as the effective ${\cal J}$-factor. With this definition, the quantity $(\sigma_A v_{\rm rel})_0$ is defined as the component of the annihilation cross section that is independent of the relative velocity.

\subsection{DM annihilation models} 
In the often-studied case in which $\sigma_A v_{\rm rel}$ does not depend on the relative velocity, eq.~\eqref{eq:Jfactor} is simply proportional to the integral of the square of the density along the line-of-sight, ${\cal J} \propto \int \rho^2 d\ell$. More generally, $\sigma_A v_{\rm rel}$ does depend on the relative velocity; in this case eq.~\eqref{eq:Jfactor} must be evaluated for the given velocity dependence. 

To account for this velocity dependence, we will make the replacement relative to the above definition and parameterize the annihilation cross section in the general form, $\sigma_A v_{\rm rel} \rightarrow \sigma_A v_{\rm rel} = (\sigma_A v_{\rm rel})_0 \, S \left(v_{\rm rel}/c\right)$, with $S \equiv \left(v_{\rm rel}/c\right)^n$. We examine the following possibilities: $n=-1$ (Sommerfeld-enhanced annihilation), $n=0$ (s-wave annihilation), $n=2$ (p-wave annihilation), and $n=4$ (d-wave annihilation). These models may be realized for different assumptions for the nature of  DM and the new physics that mediates their annihilation~\cite{Boddy:2019wfg}. 
Examining these possibilities in the context of eq.~\eqref{eq:dgammadv}, we see that the different cross section models correspond to different velocity moments of the relative velocity distribution, 
\begin{equation} 
\langle \sigma_A v_{\rm rel} \rangle(\bfx) \propto \int {\rm d}^3 \bfv_{\rm rel} P_{\bfx} (\bfv_{\rm rel})  {v}_{\rm rel}^n \equiv \mu_n(\bfx),
\label{eq:moment} 
\end{equation} 
where $\mu_n$ is the $n$-th moment of the relative velocity distribution, $P_{{\bfx}}(\bfv_{\rm rel})$. Examining eq.~\eqref{eq:moment} we may then attach a physical meaning to the velocity-averaged annihilation cross section for each of the models. In the case of the s-wave, the annihilation rate
is simply proportional to the DM density squared at a given position. For the case of Sommerfeld models, eq.~\eqref{eq:moment} is proportional to the inverse moment of the relative velocity distribution, while for the s-wave, p-wave, and d-wave models, eq.~\eqref{eq:moment} corresponds to the zeroth, 2nd, and 4th moments, respectively. 

The effective $\cal{J}$-factor in eq.~\eqref{eq:Jfactor} can then be written as
\begin{align} 
{\cal J}_s(\Psi) &= \int d \ell \int d^3 \bfv_{\rm rel} P_{\bfx} (\bfv_{\rm rel}) ~\left(\frac{{v}_{\rm rel}}{c}\right)^n~ \left[\rho (r(\ell, \Psi))\right]^2 \nonumber\\
&= \int d \ell \left[\rho (r(\ell, \Psi))\right]^2 \left(\frac{\mu_n(\bfx)}{c^n}\right). 
\label{eq:Jfactor2}
\end{align} 
Therefore, depending on the particle physics model considered, the effective $\cal{J}$-factor depends on different moments of the relative velocity distribution.

We can look at each moment more closely. In the case of the p-wave, the integral
\begin{equation} 
\mu_{2}(\bfx) \equiv \int d^3 \bfv_{\rm rel} v_{\rm rel}^2 P_{\bfx} (\bfv_{\rm rel}) 
\end{equation} 
is the square of the intrinsic relative velocity dispersion of the system at a given \bfx. This provides a measure of the disordered motion of the relative velocities about \bfx.
In the case of the d-wave model, it is useful to first define the following quantity
\begin{equation} 
\kappa(\bfx) = \frac{\int d^3 \bfv_{\rm rel} v_{\rm rel}^4 P_{\bfx} (\bfv_{\rm rel})}
{\left[ \int d^3 \bfv_{\rm rel} v_{\rm rel}^2 P_{\bfx} (\bfv_{\rm rel}) \right]^2} = \frac{\mu_4(\bfx)}{(\mu_2(\bfx))^2},
\label{eq:kurtosis}
\end{equation} 
which is motivated from the general statistical definition of kurtosis.  In the case of a Maxwell-Boltzmann distribution, we have $\kappa = 1.667$. Eq.~\eqref{eq:kurtosis} is useful because it is strongly dependent on the more extreme tails of the relative velocity distribution. For smaller $\kappa$ the components of the velocity distribution are more strongly peaked near the mean value of the respective Gaussians, while for larger $\kappa$, the velocity components are more (symmetrically) broadly distributed relative to a Gaussian. As we discuss below, this has important implications for the determination of the ${\cal J}$-factors in these models.

\section{Results} 
\label{sec:results}
We now move on to determining the ${\cal J}_s$-factors for each of the MW-like halos, under the assumptions of the different annihilation cross section models discussed above. 

\begin{figure}[t]
    \centering
    \includegraphics[width=0.495\textwidth]{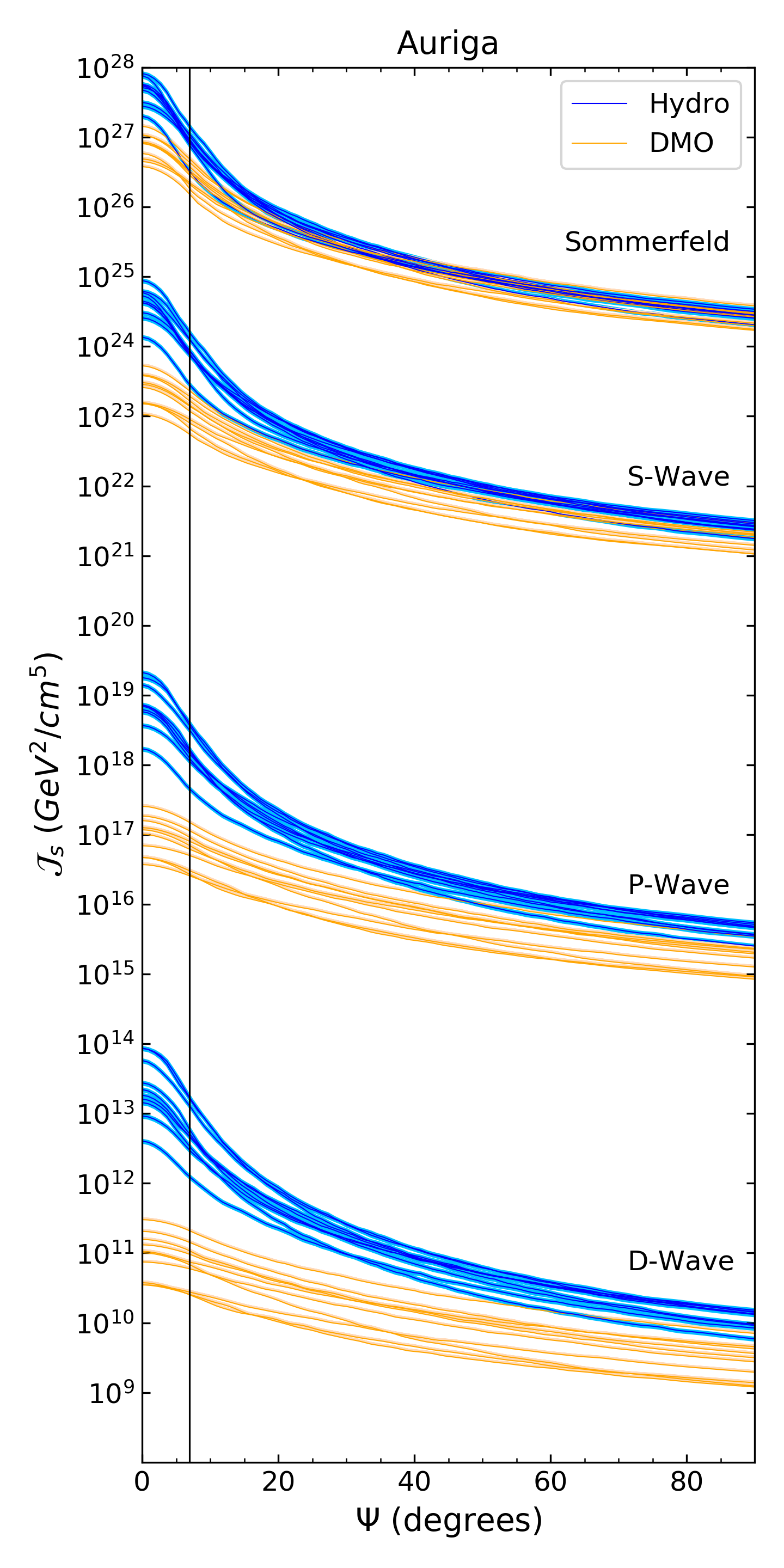}
    \includegraphics[width=0.495\textwidth]{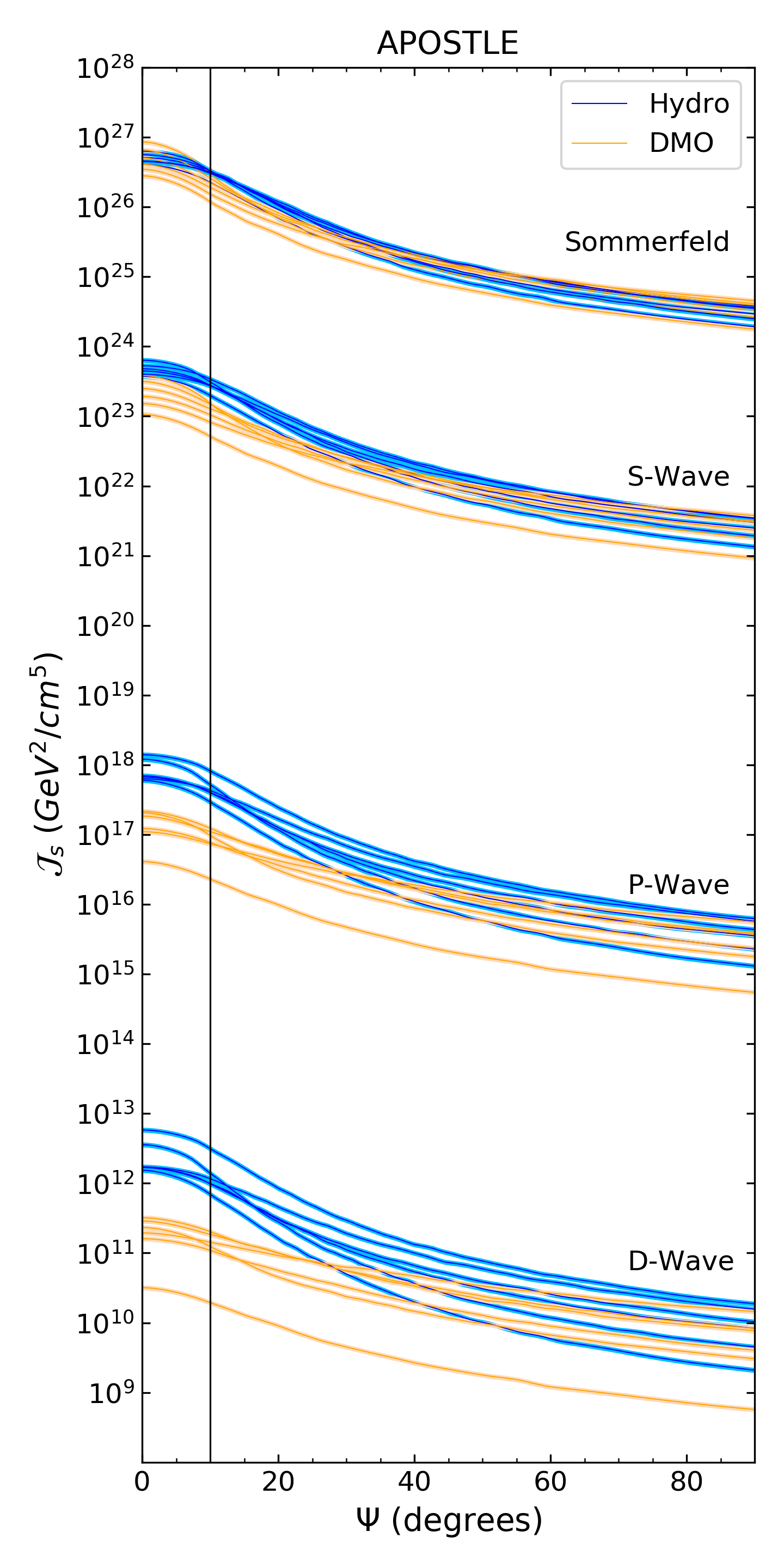}
    \caption{${\cal J}_s$-factors for the different velocity-dependent models for Auriga (left panel) and APOSTLE (right panel) simulations. For each model, we show the ${\cal J}_s$-factors for the ten MW-like halos in the hydrodynamic simulations (blue) and their DMO counterparts (yellow). The black vertical lines specify the angle $\Psi$ corresponding to the average Power radius for the Auriga and APOSTLE MW-like halos in the left and right panels, respectively.}
    \label{fig:JFactors}
\end{figure}

Figure~\ref{fig:JFactors} shows the ${\cal J}_s$-factors as a function of the angle $\Psi$ for all four cross section models for the Auriga  and APOSTLE  halos. Here we consider only the smooth halo component, so that all particles that are associated with subhalos of the main halo have been excluded. The ten Auriga MW-like halos, along with their DMO counterparts are shown in the left panel, while in the right panel we show the six APOSTLE MW-like halos and their DMO counterparts. At small angles, but still large enough to correspond to radii larger than the resolution limit, the clear trend in both simulations is for the ${\cal J}_s$-factors of the halos in the hydrodynamic simulations to be  systematically larger than those of their DMO counterparts. This behavior is primarily attributed to the contraction of the DM density profiles due to the baryons in the inner parts of the halo, as seen in figure~\ref{fig:AurigaDensityProfiles}. As discussed before, in the APOSTLE halos, the contraction of the density profiles is smaller  due to their smaller stellar masses, compared to Auriga halos. Hence, the difference between the ${\cal J}_s$-factors of the halos in the DMO and hydrodynamic simulations are also smaller.

\begin{figure}[t]
    \centering
    \includegraphics[width=0.495\textwidth]{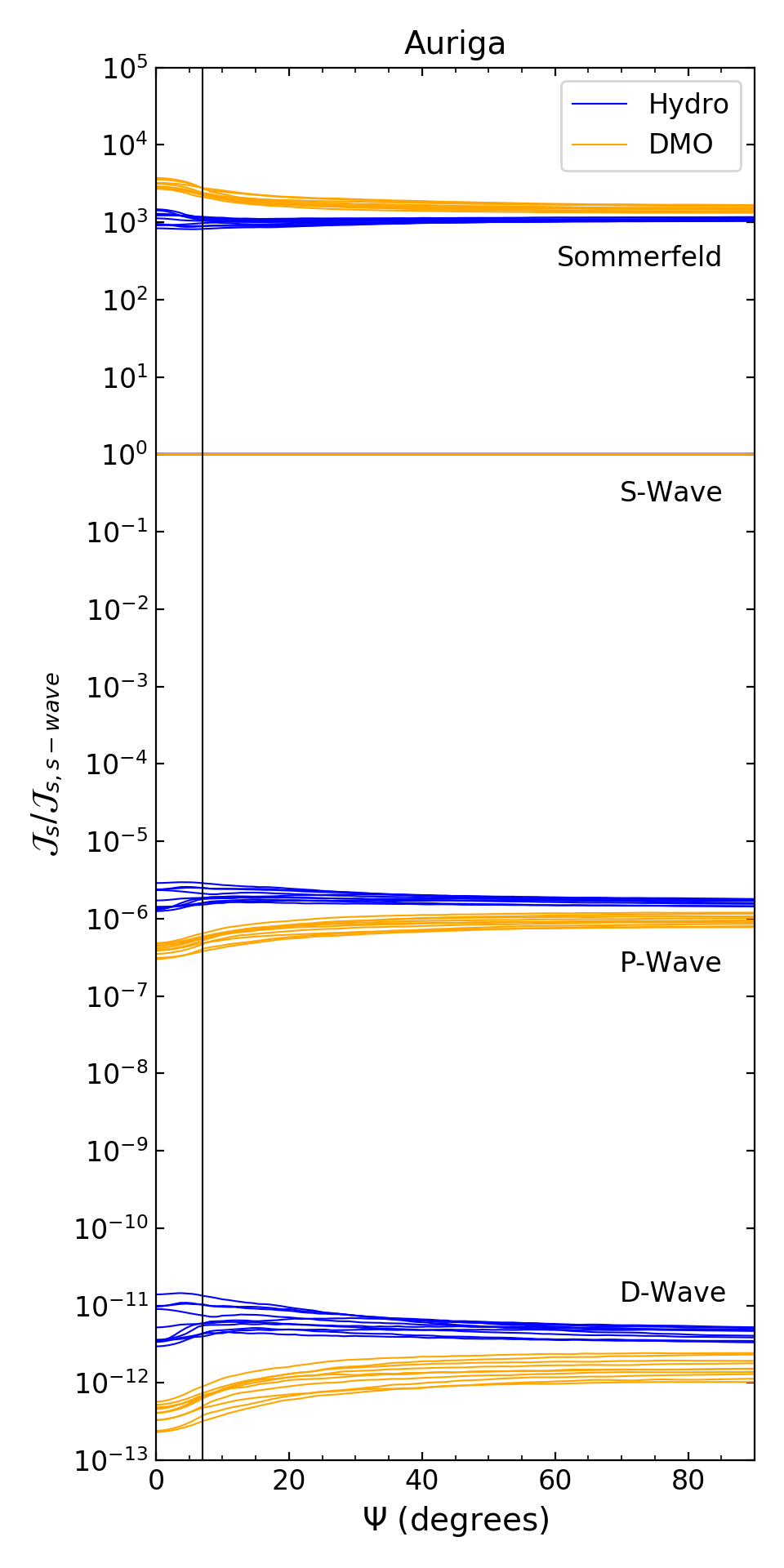}
    \includegraphics[width=0.495\textwidth]{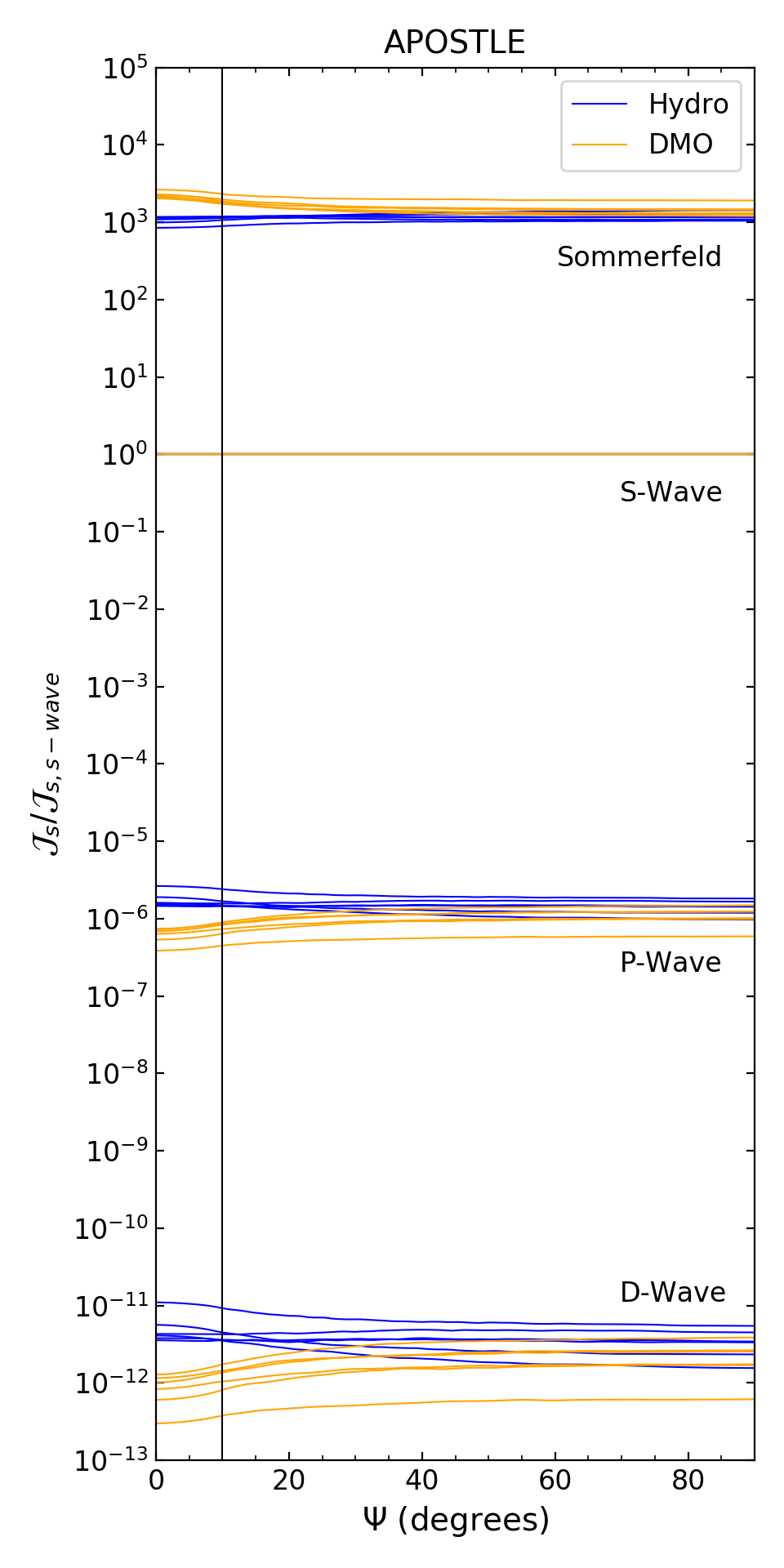}
    \caption{${\cal J}_s$-factors as in figure~\ref{fig:JFactors}, only plotted as a ratio relative to the s-wave value.}
    \label{fig:JFactor Ratios}
\end{figure}

Though the higher density of the halos in the hydrodynamic simulations at small radii provides a simple explanation for why the ${\cal J}_s$-factors are larger in the hydrodynamic case for all models, it is interesting to note the relative change in the ${\cal J}_s$-factor between the halos in the hydrodynamic simulations and their DMO counterparts for each model. Examining figure~\ref{fig:JFactors}, we see that the largest relative change occurs when going from the DMO to the hydrodynamic case for the d-wave model. On the other hand, the smallest relative change occurs for the Sommerfeld model. The larger relative increase in the ${\cal J}_s$-factor for the d-wave is a reflection of the fact that the ${\cal J}_s$-factor in this case scales as the fourth moment of the relative velocity dispersion. To appreciate quantitatively the effect of the various velocity scalings, in figure \ref{fig:JFactor Ratios}, we show the ratios of the ${\cal J}_s$-factors of each model relative to the s-wave value. 

\begin{figure}[t]
    \includegraphics[width=0.495\textwidth]{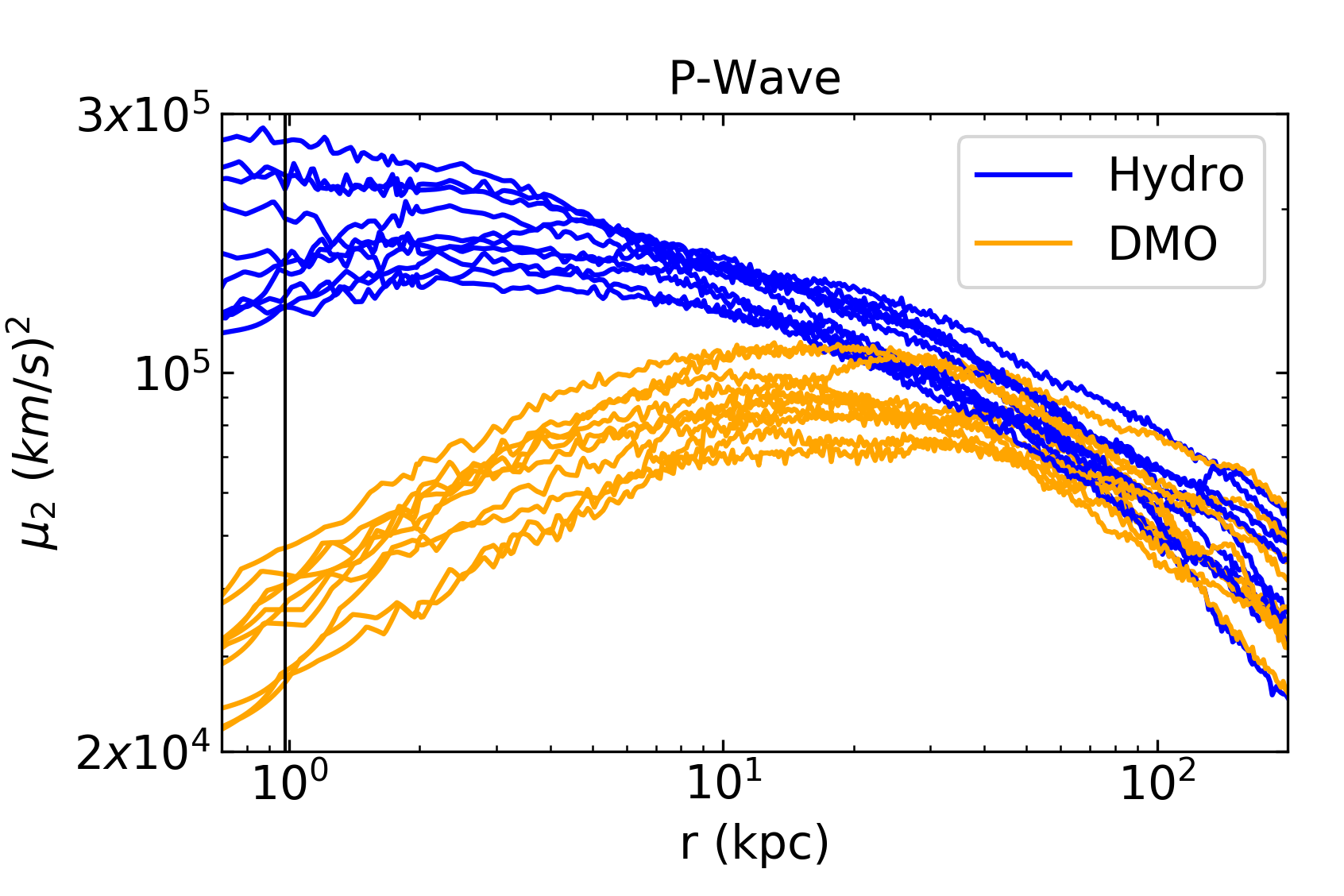}
    \includegraphics[width=0.495\textwidth]{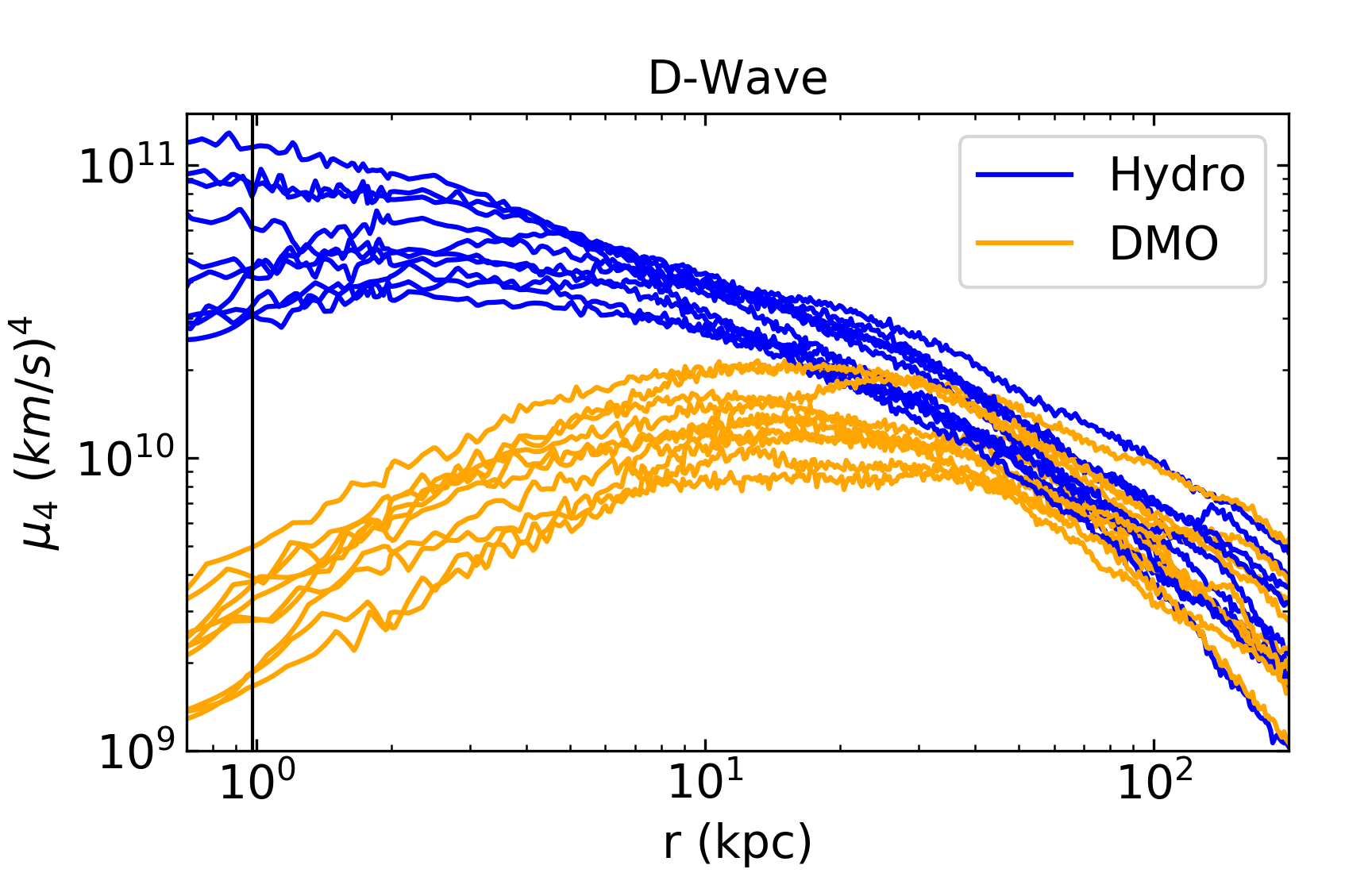}
    \includegraphics[width=0.495\textwidth]{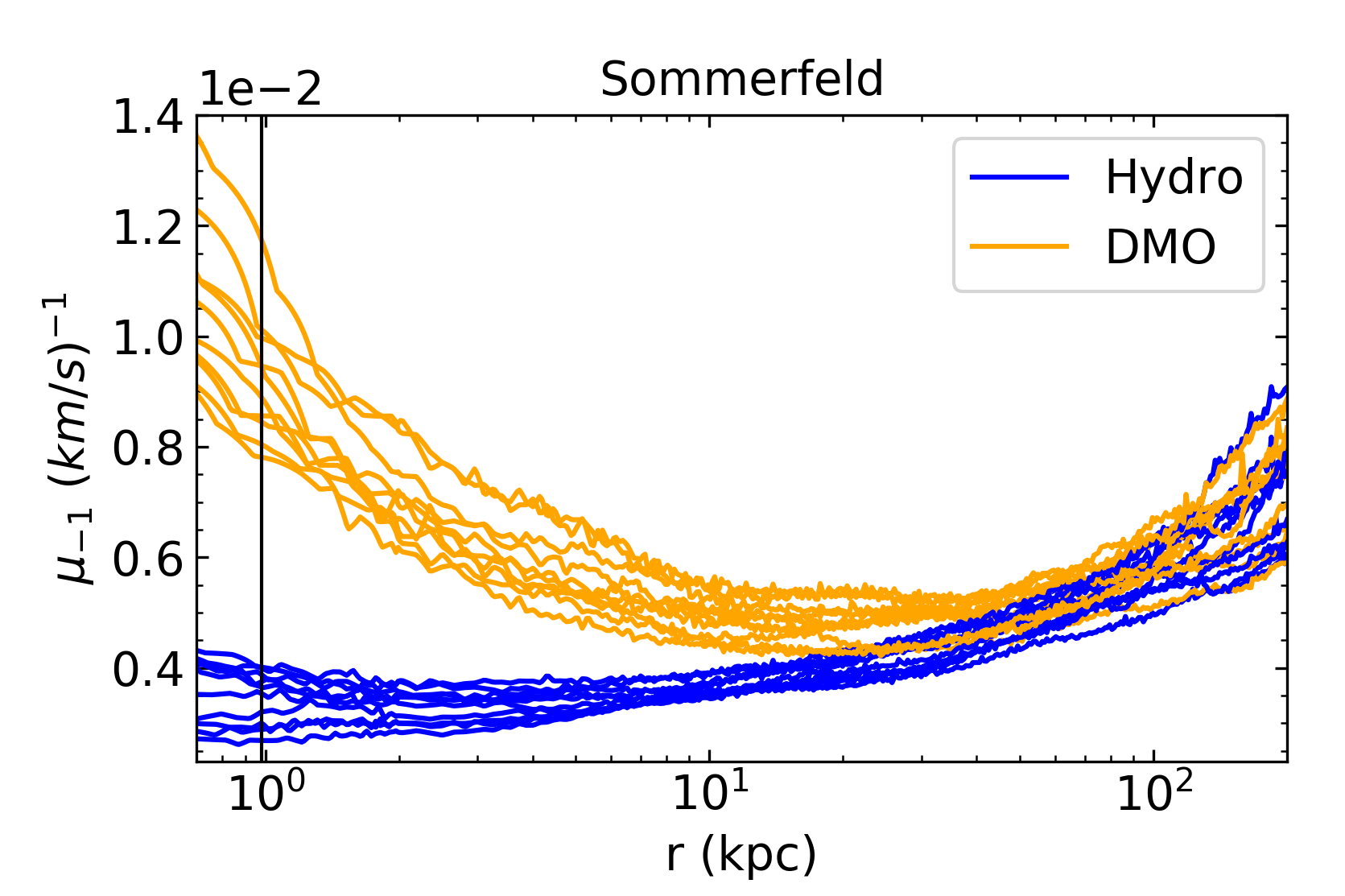}
    \includegraphics[width=0.495\textwidth]{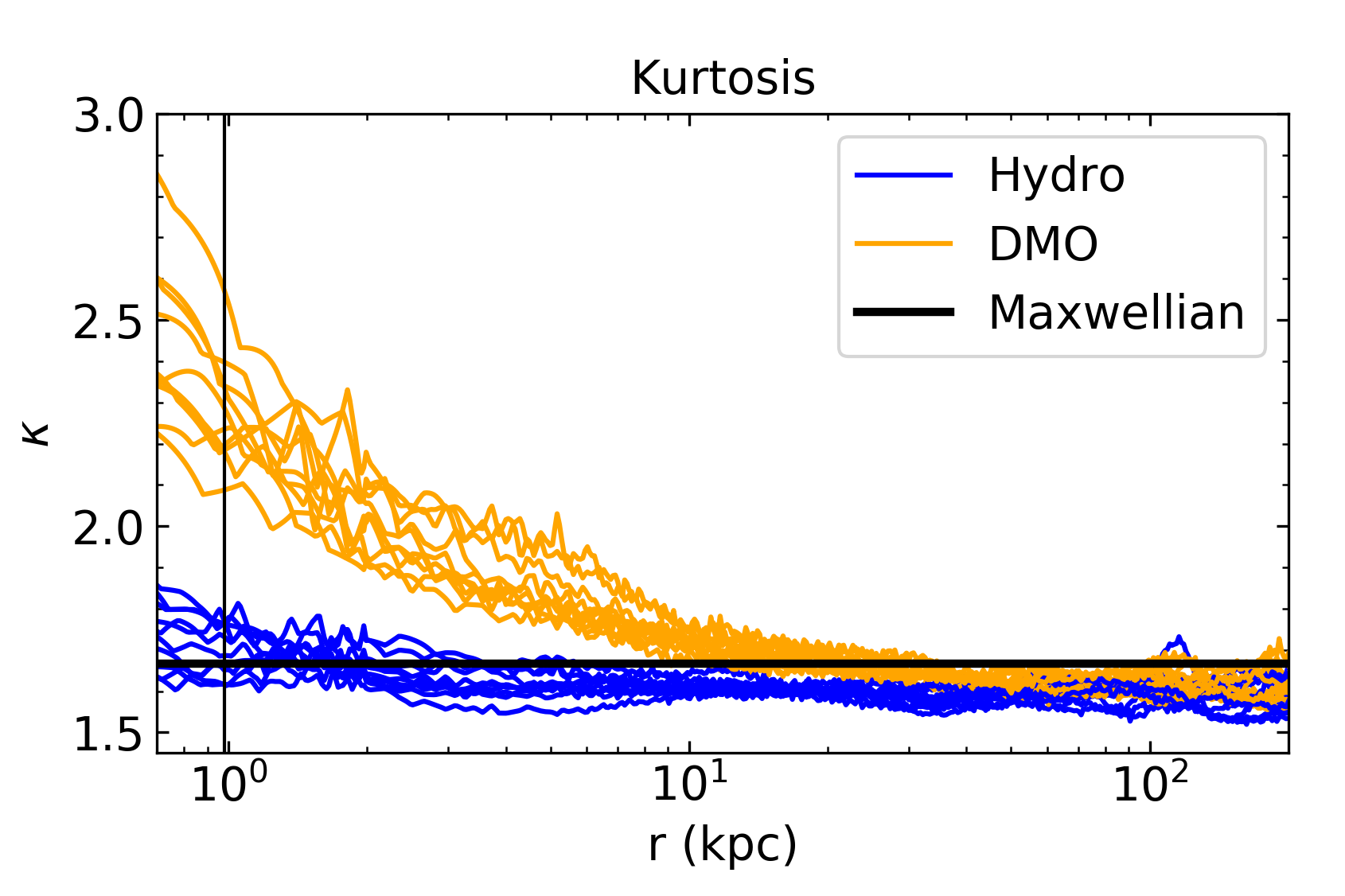}
    \caption{Velocity moments of the relative velocity distribution for the ten Auriga  MW-like halos and their DMO counterparts. The panels are: Second moment (top left), Fourth moment (top right), inverse moment (bottom left). The bottom right panel shows the fourth moment divided by the square of the second moment, with the black horizontal line indicating this quantity for the Maxwell-Boltzmann relative velocity distribution. The black vertical lines specify the average Power radius of the Auriga halos.}
    \label{fig:velocity moments}
\end{figure}

Figure~\ref{fig:velocity moments} shows the relative velocity moments for the Auriga MW-like halos, for the p-wave, d-wave and Sommerfeld models. The bottom right panel of figure~\ref{fig:velocity moments} shows the kurtosis, as defined in eq.~\eqref{eq:kurtosis}. As discussed above, the fourth moment is more sensitive to the small, but manifest differences in the tails of the relative velocity distribution as compared to a Maxwell-Boltzmann distribution. Comparing figures~\ref{fig:JFactors} and~\ref{fig:velocity moments}, we see that the scatter in the moment can be directly translated over to the scatter in the J-factor in each case.

In addition to the shift in the ${\cal J}_s$-factor itself, it is important to quantify the scatter in this quantity amongst the ten MW-like halos. Similar to the above, we find that the largest scatter is in the ${\cal J}_s$-factor of the d-wave model, and the smallest scatter is in the Sommerfeld model. In the case of the d-wave, this is again a result of the sensitivity of the ${\cal J}_s$-factor to the tails of the velocity distribution in these models. The integrand of the relative velocity moment, which in this case scales as $v_{\rm rel}^4 f(v_{\rm rel})$, exhibits a significant halo-to-halo scatter at the highest $v_{\rm rel}$, while at the lowest $v_{\rm rel}$, this integrand is nearly identical for all halos. At the other extreme for the Sommerfeld model there is significantly less scatter in the inverse moments, as shown in figure~\ref{fig:velocity moments} for the Auriga halos. In this case the integrand of the velocity moments scales as $f(v_{\rm rel})/v_{\rm rel}$, and the scatter in this integrand at the largest $v_{\rm rel}$ is much less than for the d-wave case. In addition, at low $v_{\rm rel}$, the scatter in the integrand increases, partially compensating for the scatter at high $v_{\rm rel}$. Together, these effects combine to make the halo-to-halo scatter for the Sommerfeld model the smallest amongst our cross section models. 

\begin{figure}[t]
\minipage{0.44\textwidth}
  \includegraphics[width=\linewidth]{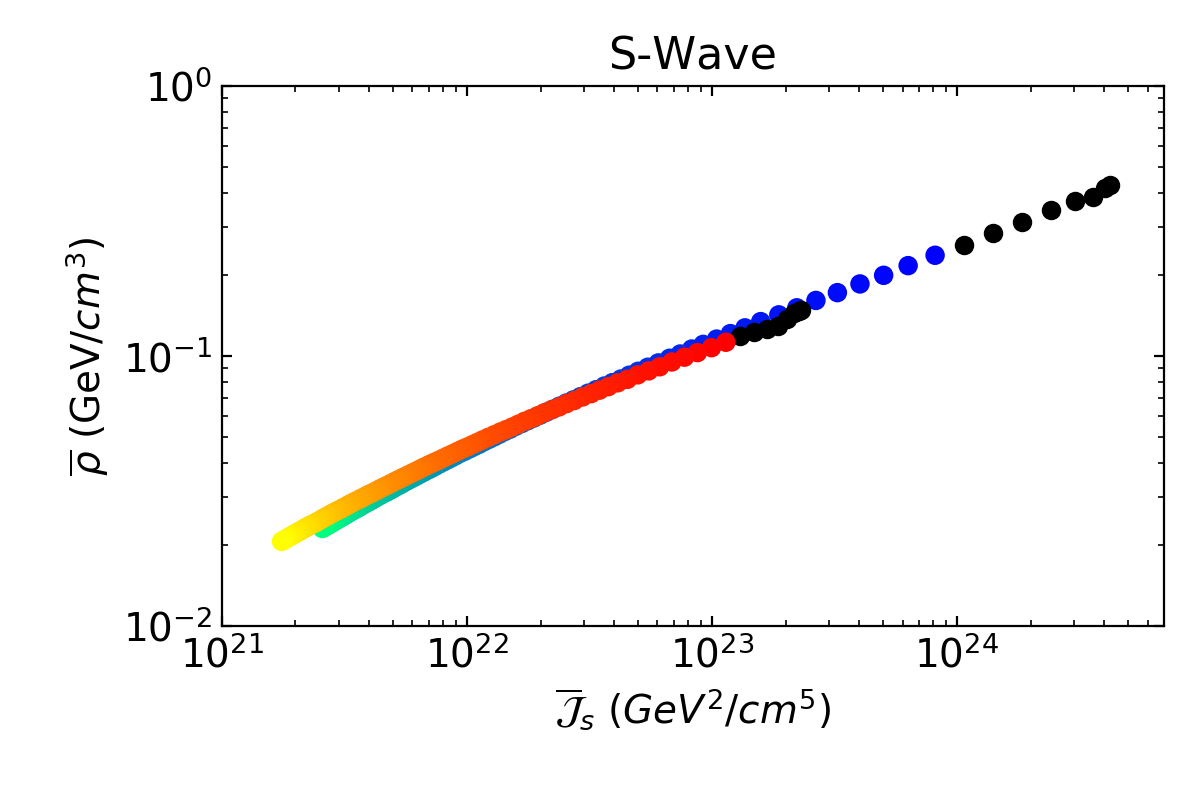}
\endminipage\hfill
\minipage{0.44\textwidth}
  \includegraphics[width=\linewidth]{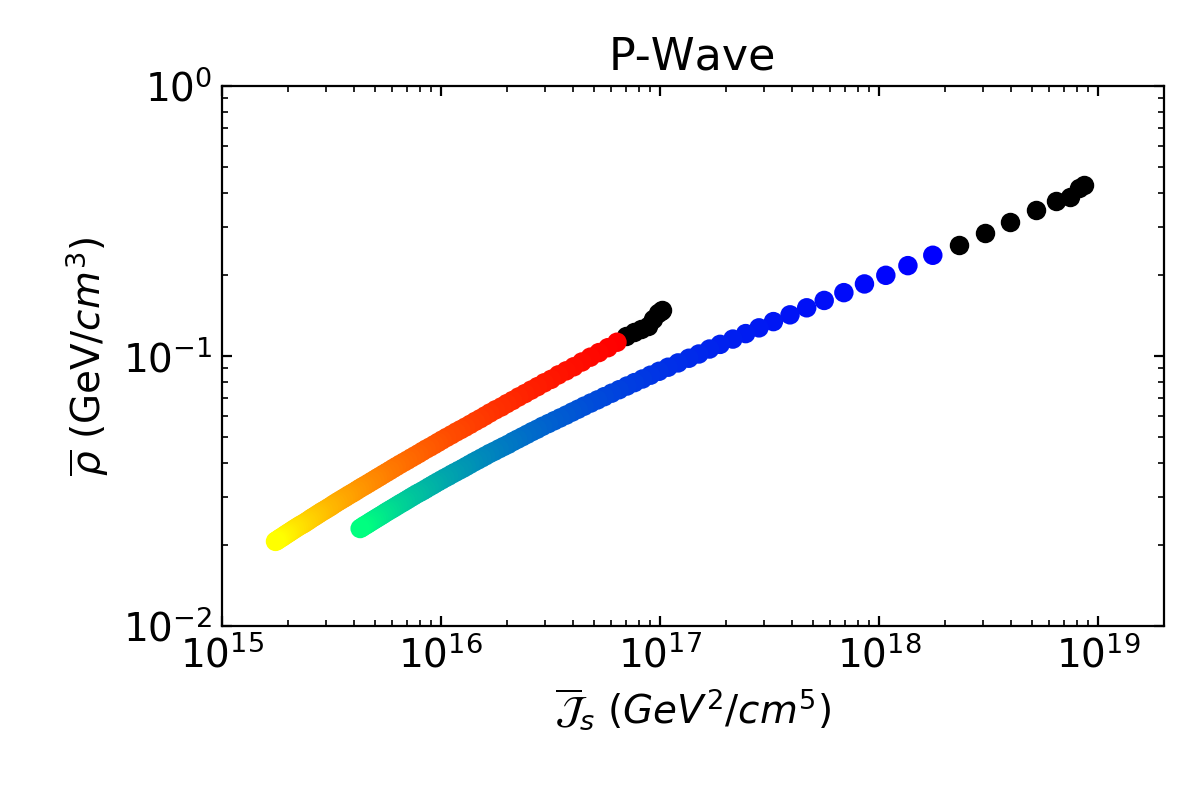}
\endminipage\hfill
\minipage{0.08\textwidth}%
  \includegraphics[width=\linewidth]{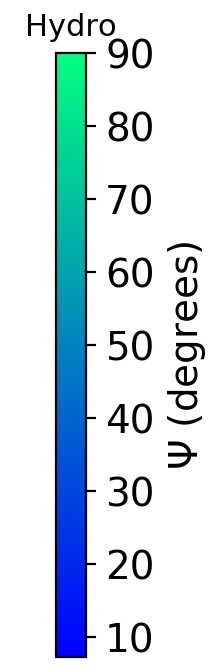}
\endminipage\hfill
\minipage{0.44\textwidth}
  \includegraphics[width=\linewidth]{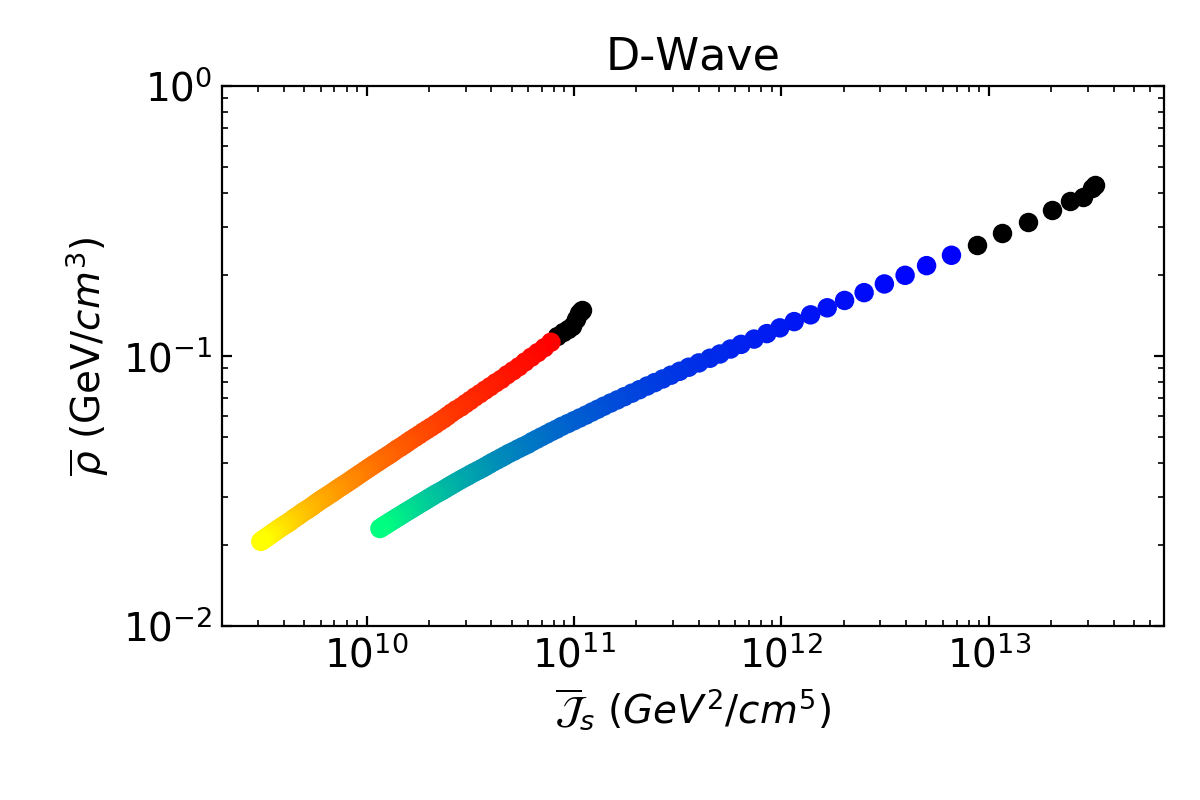}
\endminipage\hfill
\minipage{0.44\textwidth}
  \includegraphics[width=\linewidth]{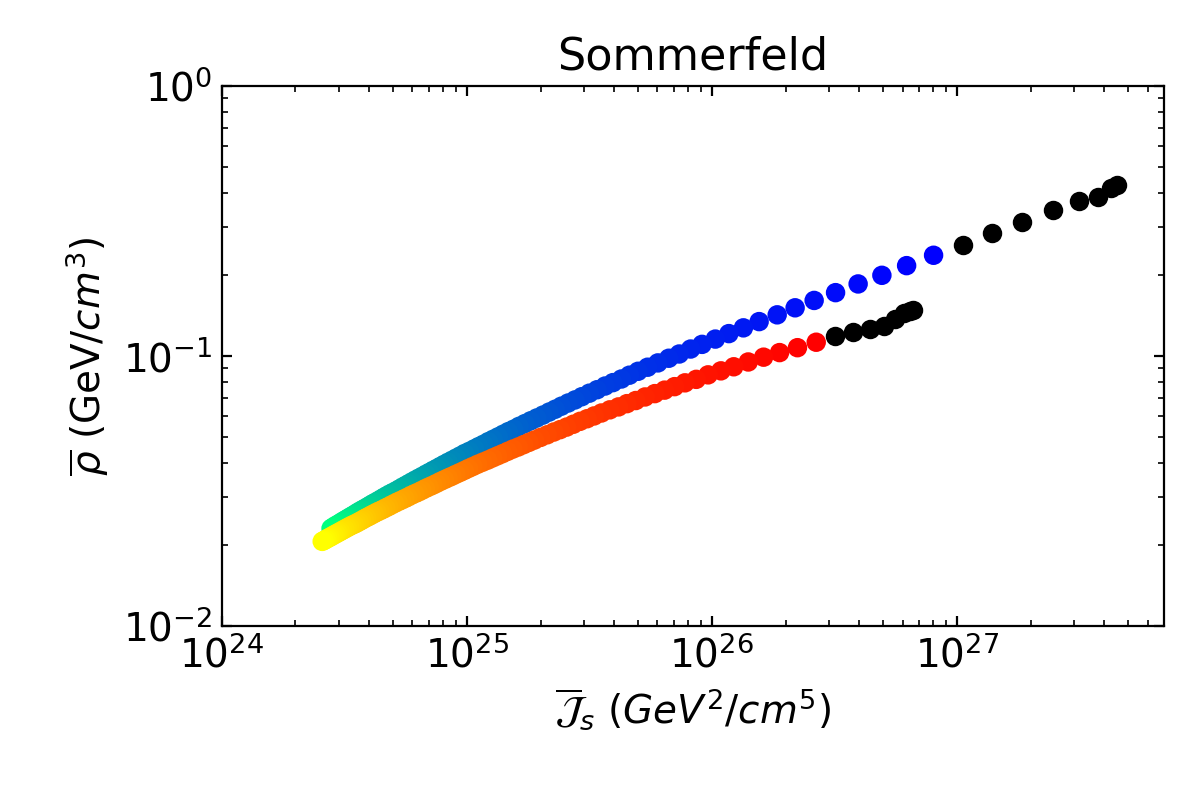}
\endminipage\hfill
\minipage{0.08\textwidth}%
  \includegraphics[width=\linewidth]{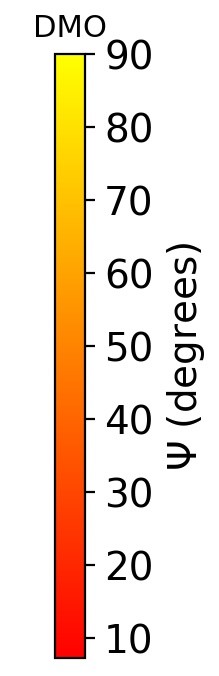}
\endminipage
\caption{Correlation between the DM density and ${\cal J}_s$-factor for Auriga halos (green to blue colored points) and their DMO counterparts (yellow to red colored points). Each point represents the average of the density and ${\cal J}_s$-factors over all the halos ($\bar{\rho}$ and $\bar{{\cal J}}_s$, respectively), along a line-of-sight at a given angle $\Psi$. The color bars on the right indicate the values of the angle from the galactic center. Angles start from $\simeq 10$ degrees, as angles at lower radii are below the resolution limit (specified by black points on the plots). Each panel shows this correlation for a different cross section model.}
    \label{fig:Correlation-rho} 
\end{figure}

\begin{figure}[!htb]
\minipage{0.44\textwidth}
  \includegraphics[width=\linewidth]{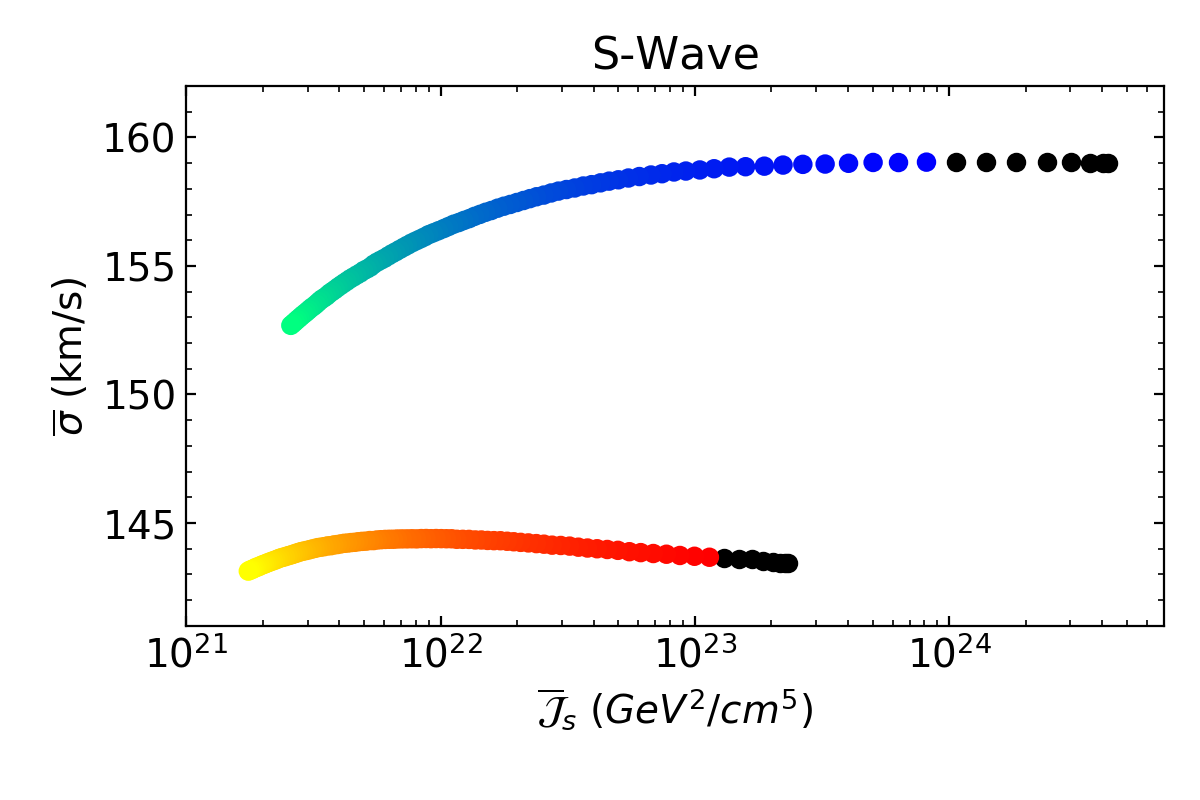}
\endminipage\hfill
\minipage{0.44\textwidth}
  \includegraphics[width=\linewidth]{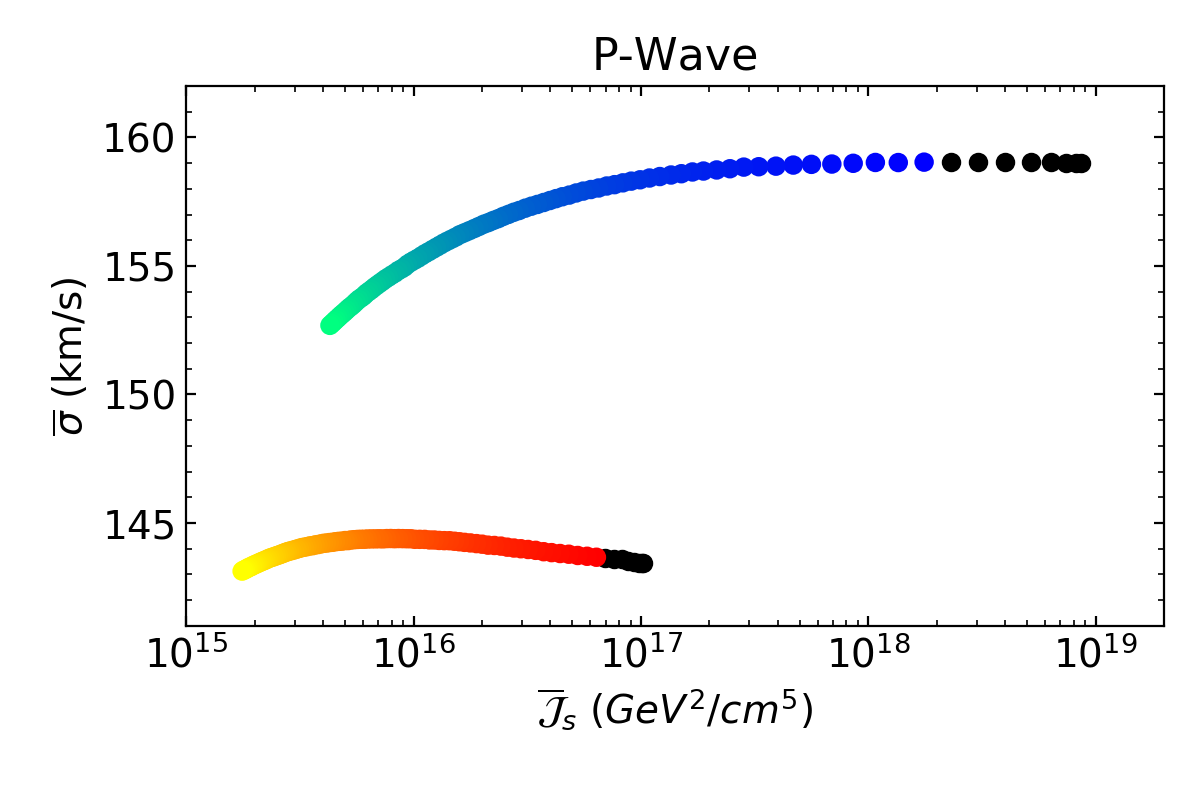}
\endminipage\hfill
\minipage{0.08\textwidth}%
  \includegraphics[width=\linewidth]{Figs/Hydro_Colorbar.png}
\endminipage\hfill
\minipage{0.44\textwidth}
  \includegraphics[width=\linewidth]{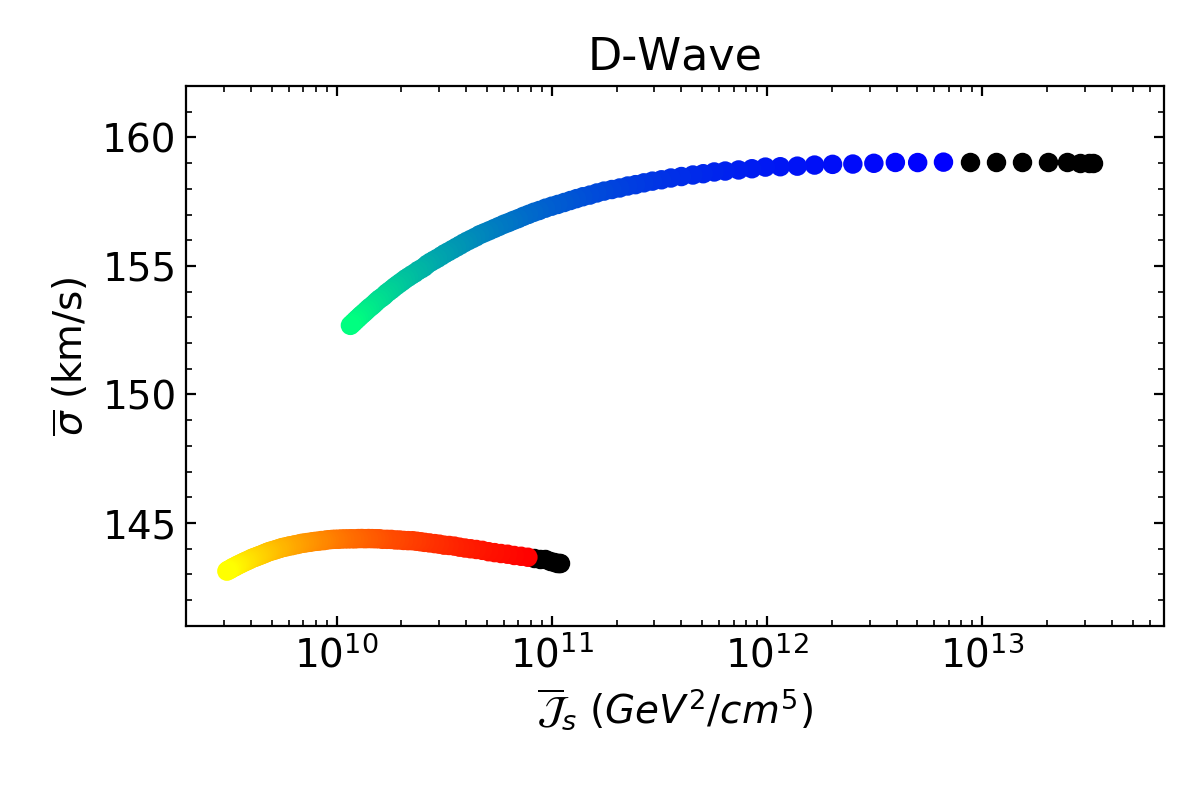}
\endminipage\hfill
\minipage{0.44\textwidth}
  \includegraphics[width=\linewidth]{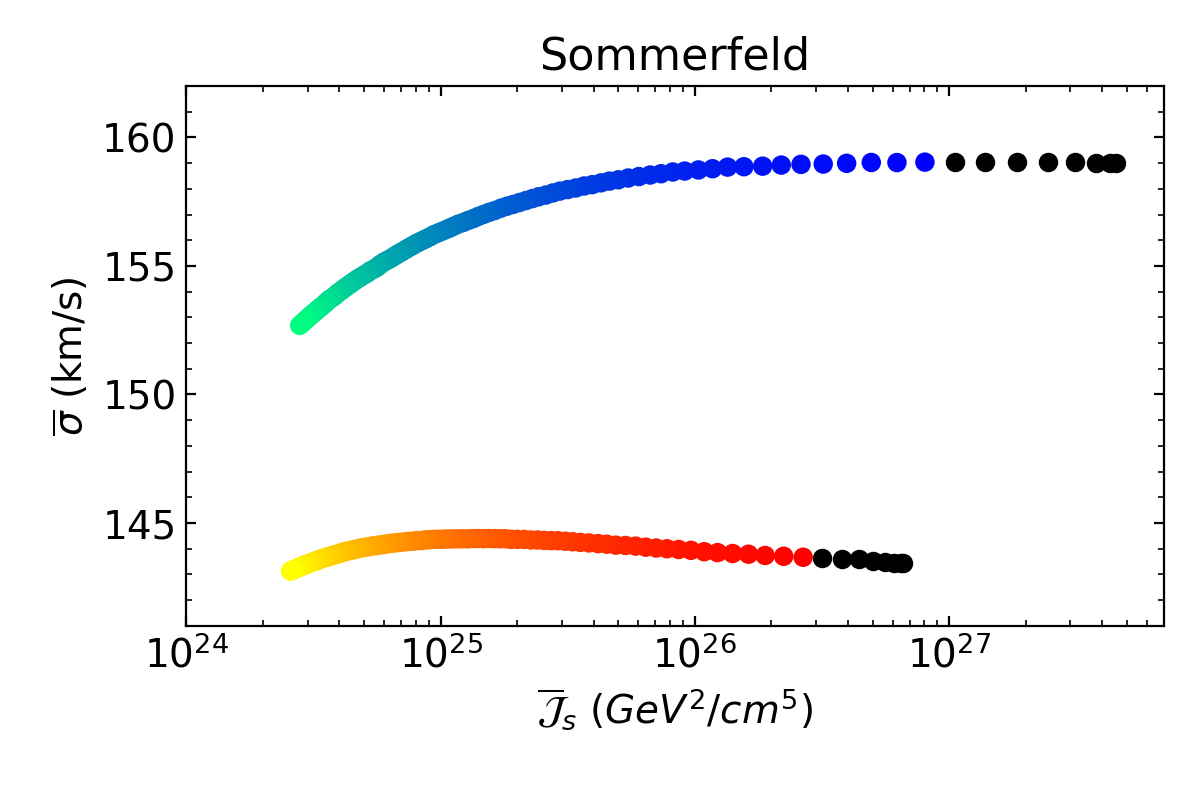}
\endminipage\hfill
\minipage{0.08\textwidth}%
  \includegraphics[width=\linewidth]{Figs/DMO_Colorbar.png}
\endminipage
\caption{Same as figure~\ref{fig:Correlation-rho}, except for the average velocity dispersion, $\bar{\sigma}$, instead of the density. Note that while figure~\ref{fig:Correlation-rho} uses log-log axes, the data in this figure is presented on semi-log axes.} 
\label{fig:Correlation-sigma}
\end{figure}

The features in the relative velocity distributions explain the relative differences between the ${\cal J}_s$-factor  of the halos in the hydrodynamic simulations and their DMO counterparts for a given annihilation cross section model. More generally, in all cases we find that the scaling of the ${\cal J}_s$-factors with angle is essentially entirely driven by the DM density profiles, and that this scaling depends very weakly on the characteristics of the DM relative velocity distributions. This can be best quantified by considering different lines-of-sight through a halo, which correspond to different values of $\Psi$, and averaging the DM density and the velocity dispersion along each line-of-sight. Figures~\ref{fig:Correlation-rho} and~\ref{fig:Correlation-sigma} show the average DM density and velocity dispersion of the ten Auriga halos, respectively, against their average ${\cal J}_s$-factor, with each point in this plane representing a different value of $\Psi$. We see from figure~\ref{fig:Correlation-rho} that for each cross section model, the average density correlates with the average ${\cal J}_s$-factor, while from figure~\ref{fig:Correlation-sigma}, there is minimal correlation with the average velocity dispersion in each case. This implies that, even for velocity dependent models, understanding the systematics in the DM density is the most important factor in determining the ${\cal J}_s$-factor.

We reiterate that the analysis of this paper has focused on determining the ${\cal J}$-factors for the smooth halo component. The contribution from DM subhalos bound to the host galaxy is expected to boost the ${\cal J}$-factor for each annihilation model. For halos in the hydrodynamic simulations and assuming s-wave annihilation, the boost factor from resolved subhalos is expected to be small, corresponding for $\lesssim 1\%$ increase over the smooth halo contribution~\cite{Grand:2020bhk}. While determining the boost factor for velocity-dependent models is beyond the scope of our current analysis, we can roughly estimate the increase in density due to subhalos by including the particles bound to subhalos\footnote{More precisely, DM particles bound to subhalos belonging to the  same \emph{friends-of-friends}~\cite{FoF-1985} group as the main halo are included, with a dimensionless linking length of 0.2 times the mean interparticle spacing.} in our calculations, and determining the spherically-averaged density and velocity distributions. With the subhalos included, we find at most $\sim 20\%$ increase in the ${\cal J}_s$-factors, which is manifest at values of $\Psi$ near the resolution limit of our simulations. This justifies our approach of focusing on the smooth halo, and indicates that the inclusion of subhalos leads to only a small increase in the ${\cal J}$-factors over the scales that we consider.

\section{Discussion and conclusions} 
\label{sec:conclusions}

In this paper we have performed the first study of the dark matter relative velocity distribution of Milky Way-like halos, using the Auriga and APOSTLE cosmological simulations. We find that the dark matter pair-wise relative velocity distribution at nearly all radii in the halos is consistent with the Maxwell-Boltzmann distribution. This agreement is particularly good for the simulations that include baryons. For the corresponding dark matter only-simulations, the agreement with the Maxwell-Boltzmann distribution is good, though there are some notable deviations, particularly at small radii as the center of the halo is approached.  

We have explored the implications for velocity-dependent dark matter annihilation, focusing on the Sommerfeld ($1/v$), s-wave ($v^0$), p-wave ($v^2$), and d-wave ($v^4$) models. We generally show that the ${\cal J}$-factors scale as the moments of the relative velocity distribution, and that the halo-to-halo scatter is largest for d-wave, and smallest for Sommerfeld models. 

Our results indicate that in velocity-dependent models, the ${\cal J}$-factor is strongly correlated with the dark matter density in the halo, and is very weakly correlated with the velocity dispersion. This implies that if the dark matter density in the Milky Way can be robustly determined, one can accurately predict the dark matter annihilation signal, without the need to identify the dark matter velocity distribution in the Galaxy. 

In calculating the ${\cal J}$-factors for velocity-dependent models, we have neglected the impact of dark matter substructure within the Milky Way-like galaxies. The effect of substructure has been explored for s-wave models in several previous studies~\cite{Springel:2008zz,Grand:2020bhk}, which indicate that the corrections for substructure are small, at least at the resolution limits of present simulations. It is possible that boost factors can be significant for extrapolations down to $\sim$ Earth-mass subhalos, in particular for Sommerfeld-enhanced models. Accurately calculating the boost factors for velocity-dependent models required determining the concentration-mass relation for subhalos~\cite{Wang:2019ftp} and their velocity distribution, and understanding how to extrapolate these beyond the resolution limit of the simulations. We leave this topic as a subject for future study. 

The results we have presented will be important in guiding searches for velocity-dependent dark matter annihilation, for example with Fermi-LAT data or with future data from higher-energy gamma-ray instruments. Though p-wave and d-wave annihilation may be realized in simple models~\cite{Kumar:2013iva,Giacchino:2013bta,Han:2015cty}, due to the sensitivity of these instruments, for the simplest models bounds on p-wave~\cite{Diamanti:2013bia,Boddy:2019wfg} and d-wave~\cite{Boddy:2019wfg} cross sections are much larger than those for thermal relic dark matter. Bounds may be improved upon by considering more unique astrophysical environments, for example the supermassive black hole at the center of the Milky Way~\cite{Johnson:2019hsm}. The phenomenology becomes richer for multi-state dark matter, such that Sommerfeld boosts can enhance the p-wave component and suppress the s-wave component~\cite{Das:2016ced}. The results we have presented provide the most realistic approach available to providing robust constraints on these velocity-dependent models with astrophysical systematics incorporated.

\subsection*{Acknowledgements}
We thank James Bullock, Basudep Dasgupta, Francesc Ferrer, and Jason Kumar for discussions on this paper. NB acknowledges the support of the Natural Sciences and Engineering Research Council of Canada (NSERC), funding reference number RGPIN-2020-07138. EB and LES acknowledge support from DOE Grant de-sc0010813. AF is supported by the Leverhulme Trust and the Science and Technology Facilities Council (STFC) [grant numbers ST/P000541/1]. CSF acknowledges support from the European Research Council through ERC Advanced Investigator grant, DMIDAS [GA 786910], and from the UK STFC [grant number ST/F001166/1, ST/I00162X/1, ST/P000541/1]. FM acknowledges support through the program ``Rita Levi Montalcini'' of the Italian MUR. KAO acknowledges support from the European Research Council through ERC Advanced Investigator grant, DMIDAS [GA 786910]. This work used the DiRAC Memory Intensive system at Durham University, operated by ICC on behalf of the STFC DiRAC HPC Facility (www.dirac.ac.uk). This equipment was funded by BIS National E-infrastructure capital grant ST/K00042X/1, STFC capital grant ST/H008519/1, and STFC DiRAC Operations grant ST/K003267/1 and Durham University. DiRAC is part of the National E-Infrastructure.


\clearpage

\appendix

\section{Best fit parameters for relative velocity distributions}
\label{app:fit}

In table~\ref{tab:BFMax} we present the best fit peak speeds and reduced $\chi^2$ values for the Maxwellian functional form to fit the DM relative velocity modulus distributions of the  Auriga and APOSTLE MW-like halos. The best fit parameters are given for the DM particles in spherical shells at different radii from the center of the halo.

 \begin{table}[h]
    \centering
    \begin{tabular}{|c|c|c|c|c|c|c|c|c|}
      \hline
       & \multicolumn{2}{|c|}{$r=2$~kpc} & \multicolumn{2}{|c|}{$r=8$~kpc} & \multicolumn{2}{|c|}{$r=20$~kpc} 
       &  \multicolumn{2}{|c|}{$r=50$~kpc}\\
      \hline
       Halo Name  & $v_0$~[km~s$^{-1}$] & $\chi^2_{\rm red}$ & $v_0$~[km~s$^{-1}$] & $\chi^2_{\rm red}$ & $v_0$~[km~s$^{-1}$] & $\chi^2_{\rm red}$ & $v_0$~[km~s$^{-1}$] & $\chi^2_{\rm red}$\\
       \hline
       Au2 & 315.93 & 1.05 & 321.03 & 1.99 & 307.43 & 0.91 & 263.53 & 1.33 \\
       Au4 & 337.43 & 0.98 & 335.43 & 0.86 & 297.53 & 1.05 & 245.12 & 2.09 \\
       Au5 & 379.34 & 0.46 & 338.43 & 0.63 & 293.63 & 0.96 & 236.62 & 1.14 \\
       Au7 & 308.93 & 0.68 & 298.33 & 0.72 & 268.13 & 0.80 & 225.42 & 1.24 \\
       Au9 & 384.14 & 0.51 & 328.33 & 0.71 & 274.93 & 0.72 & 226.62 & 2.12 \\
       Au12 & 341.83 & 0.56 & 314.93 & 0.81 & 273.53 & 1.01 & 235.42 & 2.12 \\
       Au19 & 326.23 & 0.62 & 299.83 & 0.63 & 280.13 & 0.86 & 233.02 & 1.88 \\
       Au21 & 331.93 & 0.21 & 330.73 & 0.63 & 303.73 & 1.28 & 246.92 & 1.67 \\
       Au22 & 401.64 & 0.61 & 316.73 & 1.64 & 270.53 & 2.48 & 220.92 & 1.47 \\
       Au24 & 363.04 & 0.40 & 329.03 & 0.53 & 302.13 & 1.13 & 249.42 & 1.63 \\
      \hline
      AP-V1-1-L2 & 309.58 & 0.89 & 312.75 & 0.49 & 299.72 & 0.73 & 267.15 & 1.55 \\
      AP-V6-1-L2 & 368.60 & 0.62 & 331.61 & 0.53 & 308.73 & 0.52 & 273.38 & 1.07 \\
      AP-S4-1-L2 & 297.77 & 0.62 & 295.07 & 0.42 & 271.10 & 0.64 & 243.93 & 0.92 \\
      AP-V4-1-L2 & 296.83 & 0.68 & 296.91 & 0.55 & 269.67 & 0.61 & 238.93 & 1.84 \\
      AP-V4-2-L2 & 298.68 & 0.92 & 244.91 & 0.90 & 229.59 & 0.66 & 198.06 & 1.17 \\
      AP-S6-1-L2 & 313.43 & 0.91 & 267.60 & 0.97 & 241.73 & 1.24 & 201.78 & 1.19 \\
      \hline
    \end{tabular}
\caption{Best fit peak speed, $v_0$, and the reduced $\chi^2$ values for the goodness of fit of the Maxwellian velocity distributions to the DM speed distributions of the  Auriga and APOSTLE MW-like halos at different radii from the center of the halo.}
\label{tab:BFMax}
 \end{table}

 \section{Components of the relative velocity distributions}
 \label{app:components}
 
\begin{figure}[t]
\begin{center}
    \includegraphics[width=0.31\textwidth]{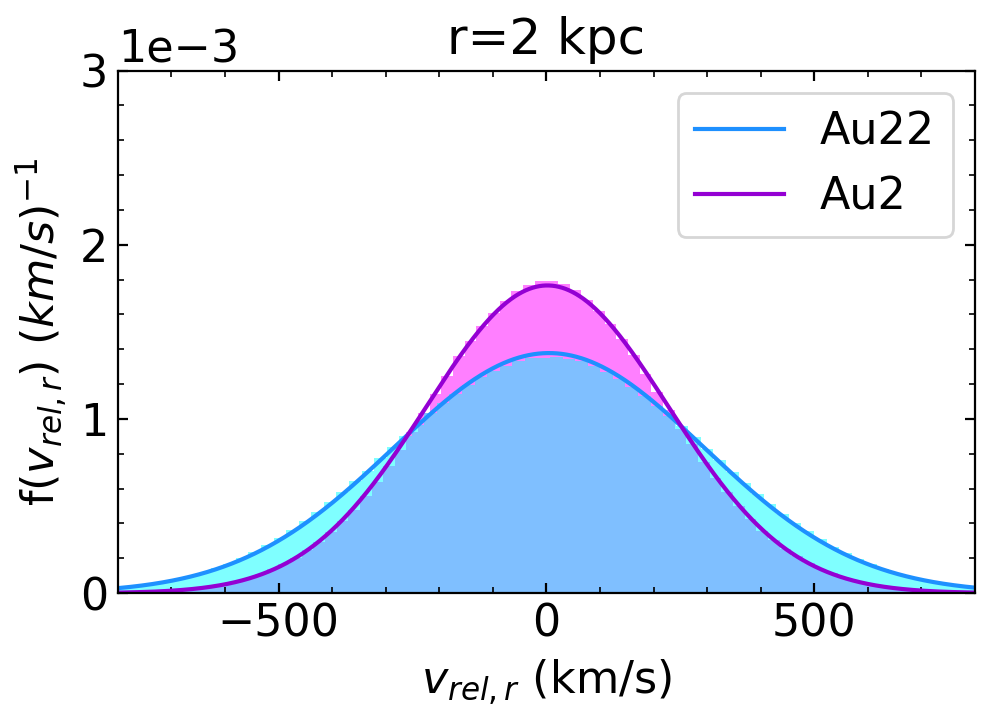}
    \vspace{0.02\textwidth}
    \includegraphics[width=0.31\textwidth]{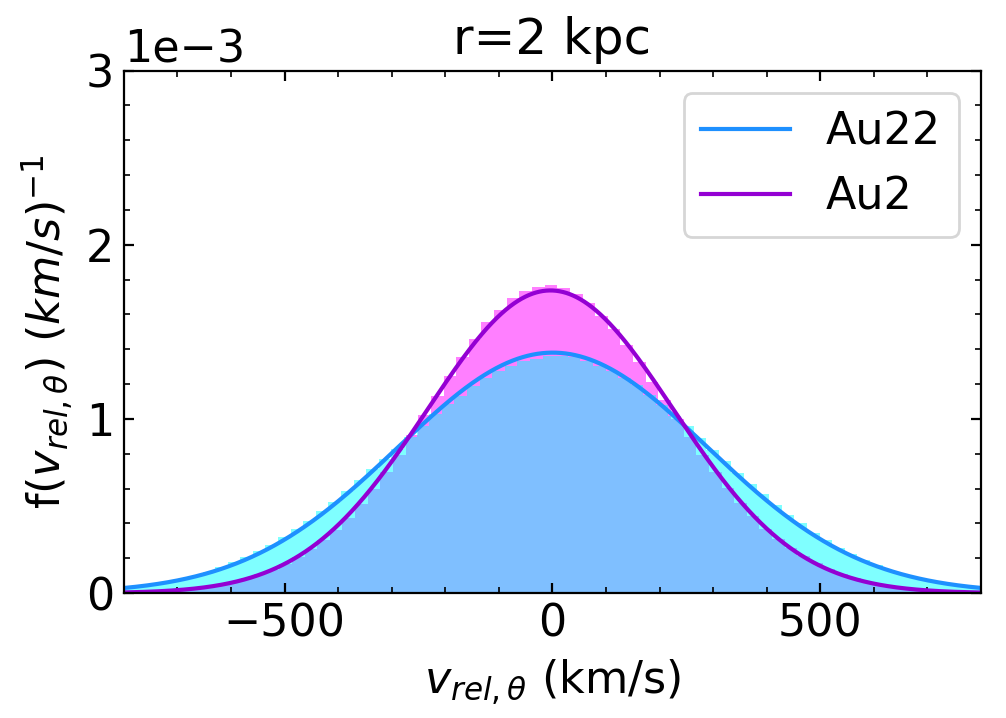}
    \includegraphics[width=0.31\textwidth]{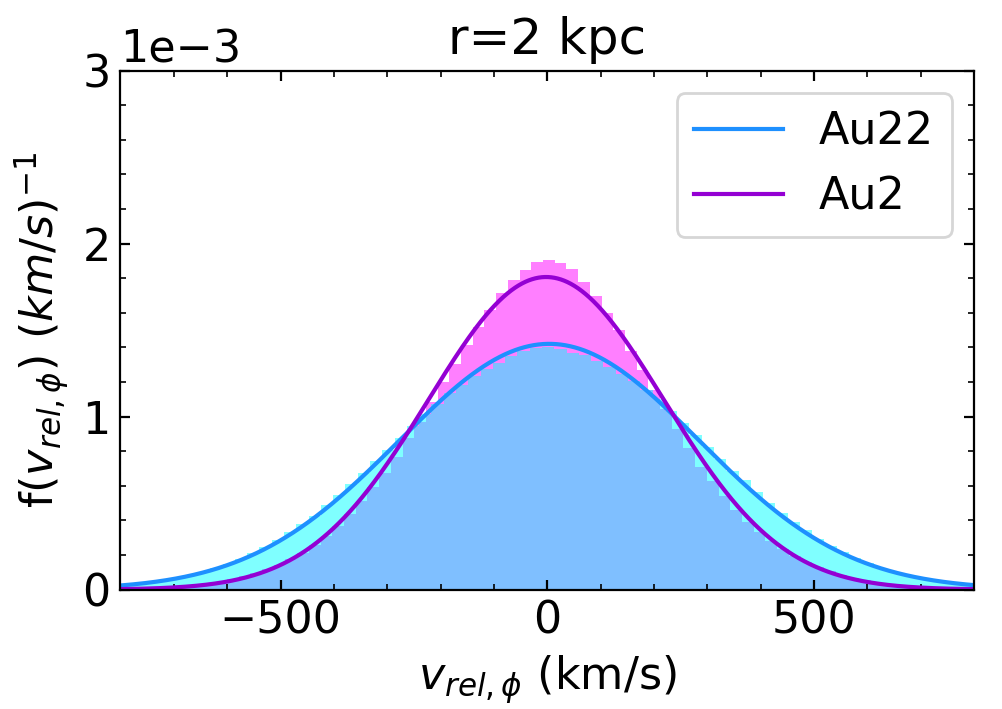}
    \vspace{0.02\textwidth}
    \includegraphics[width=0.31\textwidth]{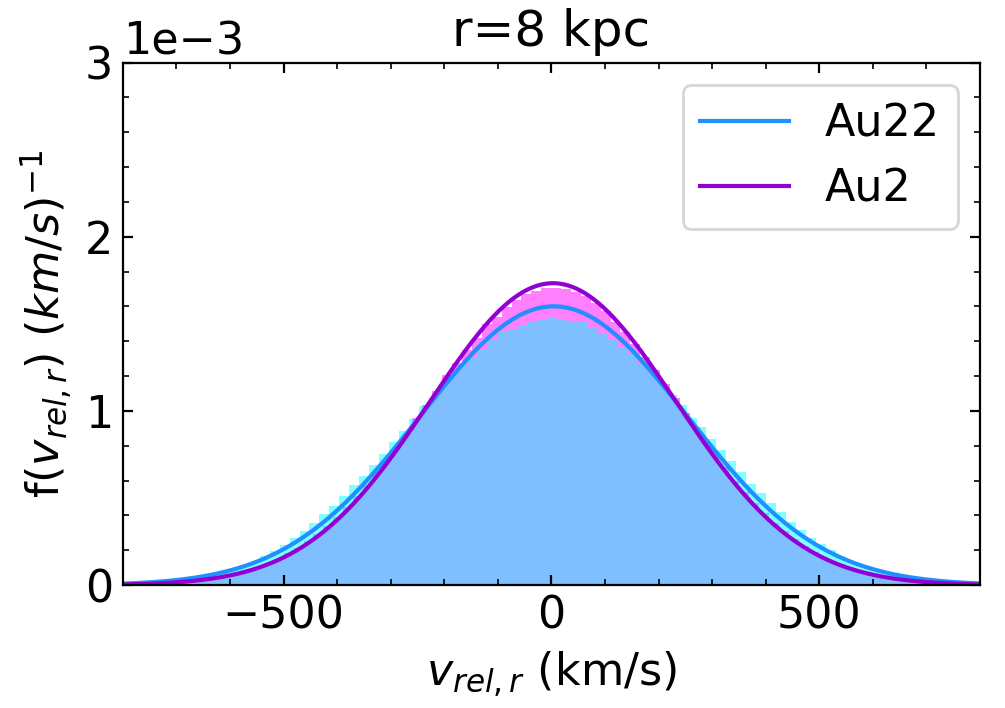}
    \includegraphics[width=0.31\textwidth]{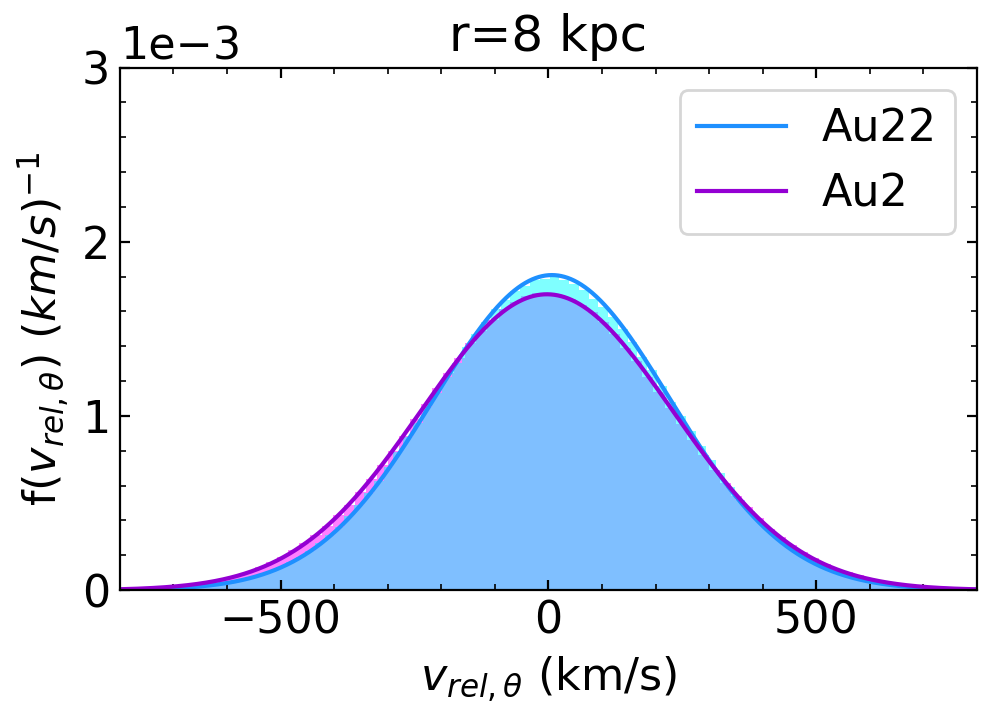}
    \includegraphics[width=0.31\textwidth]{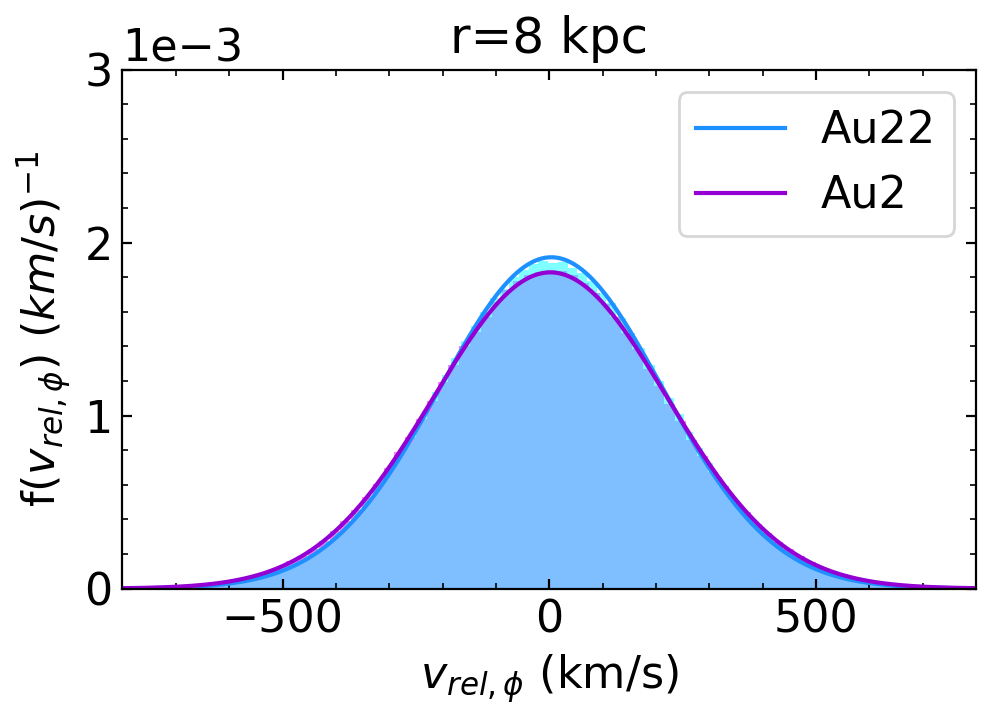}
    \vspace{0.02\textwidth}
    \includegraphics[width=0.31\textwidth]{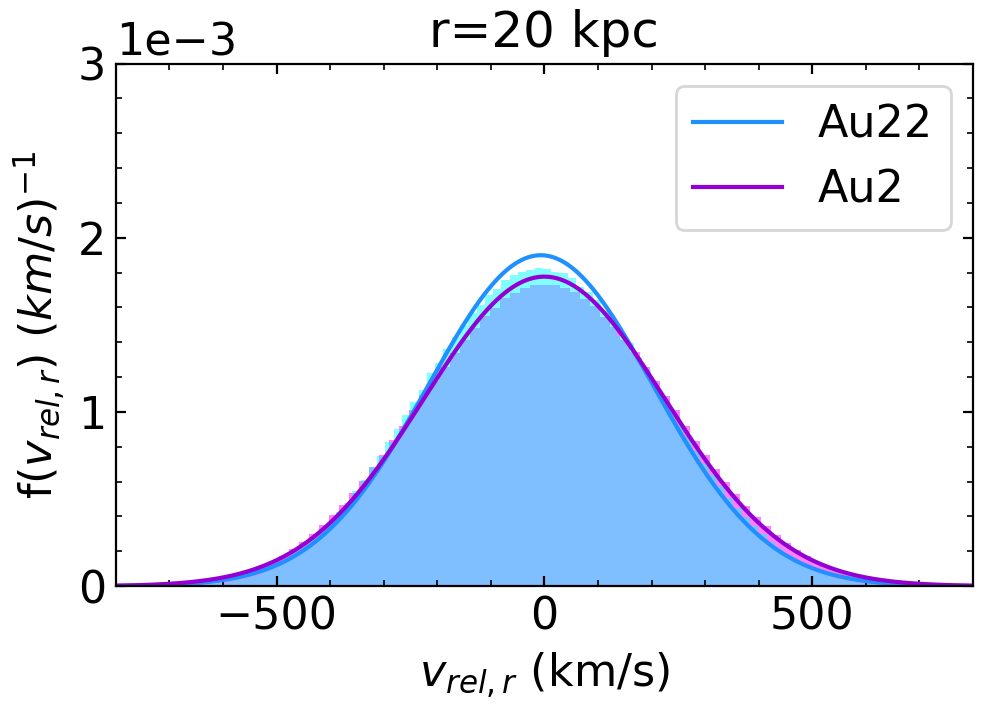}
    \includegraphics[width=0.31\textwidth]{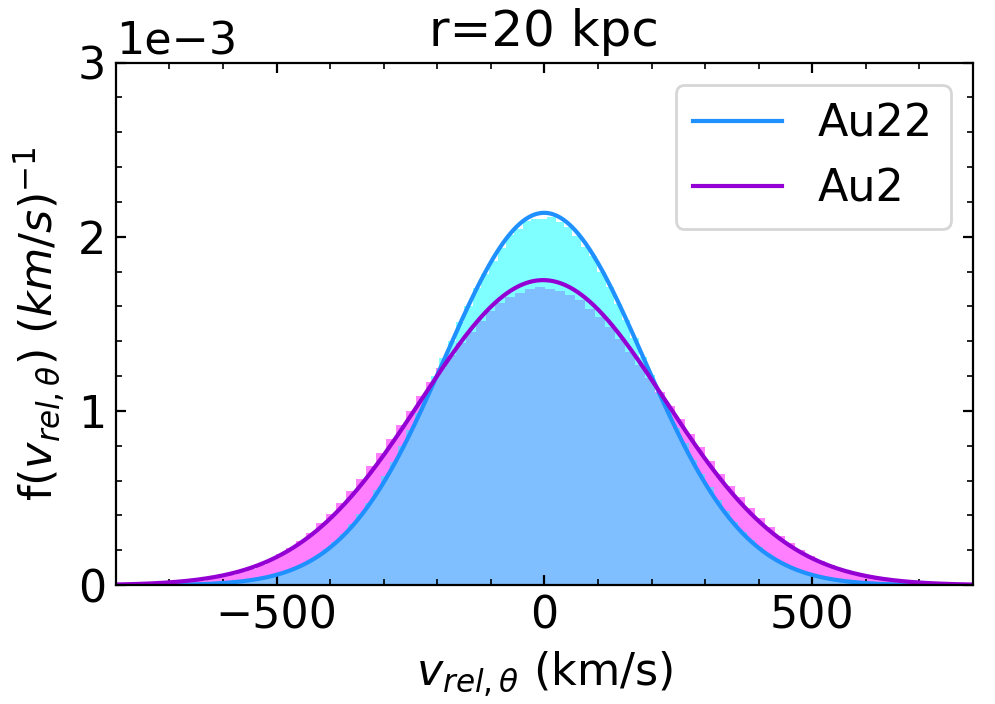}
    \includegraphics[width=0.31\textwidth]{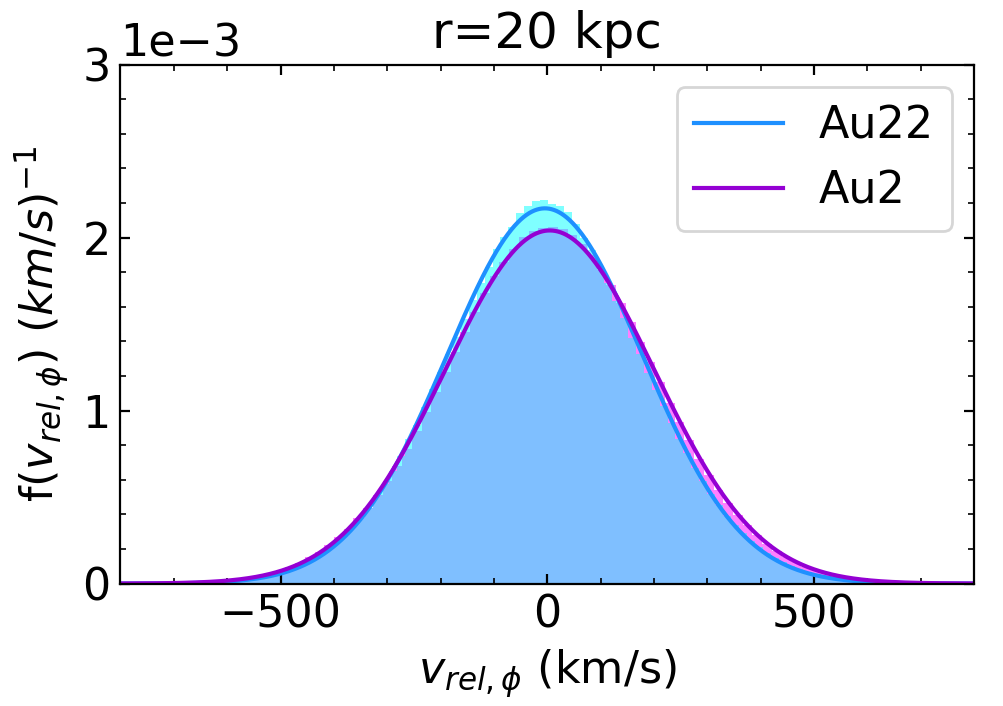}
    \includegraphics[width=0.31\textwidth]{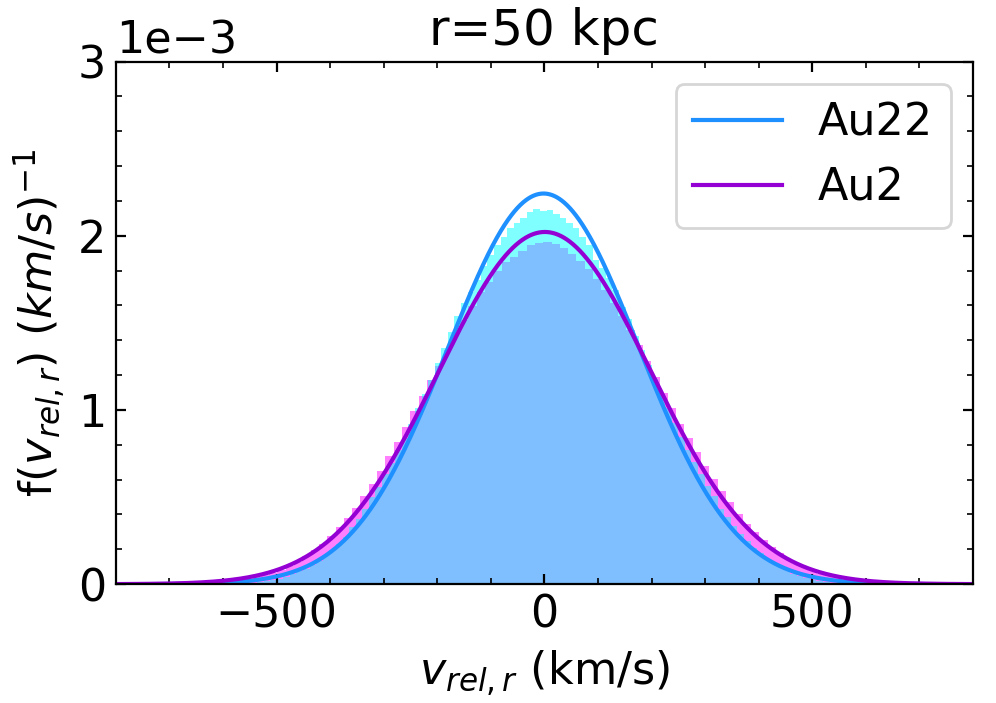}
    \includegraphics[width=0.31\textwidth]{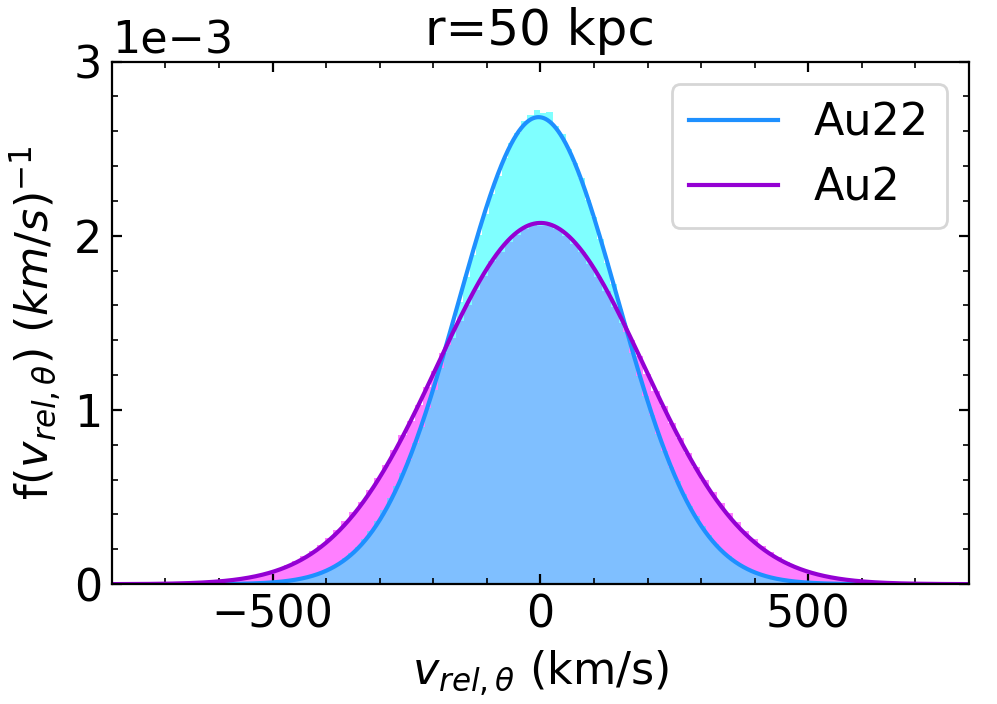}
    \includegraphics[width=0.31\textwidth]{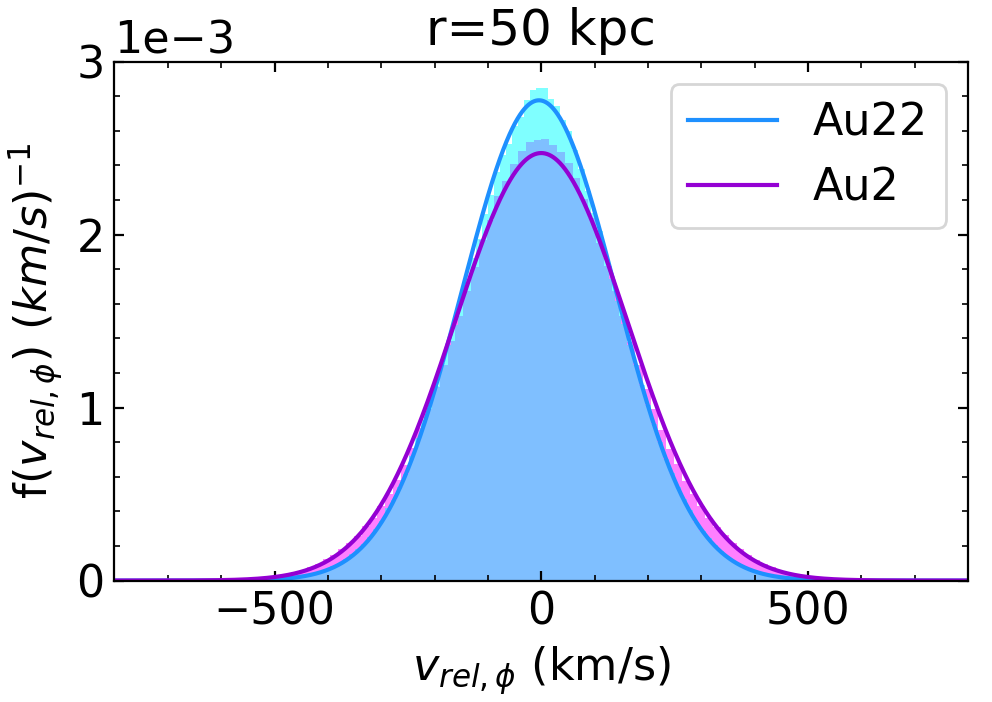}
    \caption{The histograms show the radial (left panels), polar (middle panels) and azimuthal (right panels) components of the DM relative velocity distributions for halos Au2 (magenta) and Au22 (blue). From top to bottom the rows show the distributions in radial shells at 2 kpc, 8 kpc, 20 kpc, and 50 kpc from the Galactic center. The solid lines specify the best fit Gaussian distribution for each velocity component and each halo.
}
    \label{fig:hydro vrel components}
\end{center}
\end{figure}

 In figure~\ref{fig:hydro vrel components} we show the radial ($v_{{\rm rel}, r}$), polar ($v_{{\rm rel}, \theta}$), and azimuthal ($v_{{\rm rel}, \phi}$) components of the DM relative velocity distributions for halos Au2 and Au22 at four different Galactocentric radii. The origin of our reference frame is at the Galactic center, and the $z$-axis is perpendicular to the stellar disk. The three components of the relative velocity distribution are individually normalized to unity, such that $\int dv_{{\rm rel}, i} f(v_{{\rm rel}, i})=1$ for $i=r, \theta, \phi$.

The three components of the relative velocity distribution are different at each radius, and there is a clear velocity anisotropy at all radii. The solid colored curves in each panel specify the best fit Gaussian distribution to each relative velocity component for the two halos. 

\begin{figure}[t]
\begin{center}
    \includegraphics[width=0.7\textwidth]{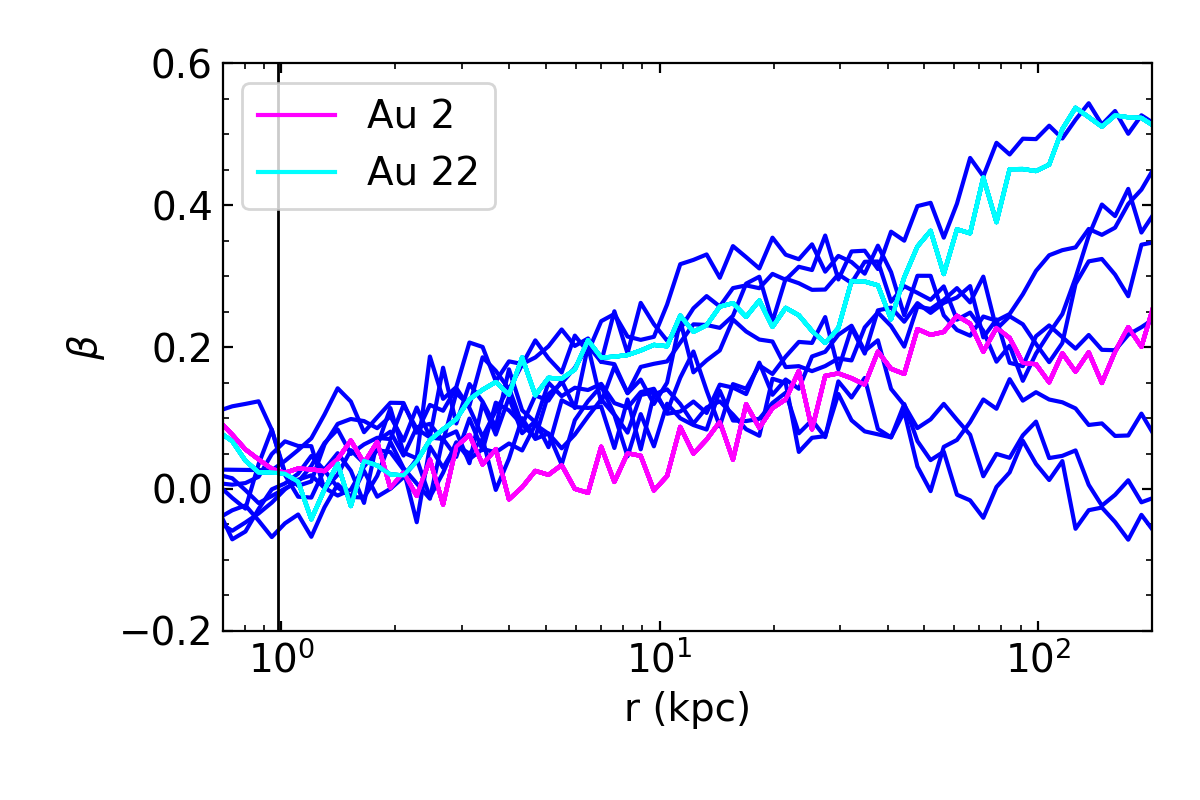}
    \caption{Anisotropy parameter, $\beta$, as a function of Galactocentric radius for the 10 Auriga MW-like halos. The cyan and magenta curves specify the anisotropy parameter for halos Au2 and Au22, respectively.}
    \label{fig:beta}
\end{center}
\end{figure}

To better understand the degree of anisotropy in the relative velocities, we compute the anisotropy parameter,
\begin{equation}
\beta = 1 - \frac{\sigma_\theta^2+\sigma_\phi^2}{2\sigma_r^2},
\end{equation}
where $\sigma_r$, $\sigma_\theta$, and $\sigma_\phi$ are the radial, polar, and azimuthal velocity dispersions, respectively. Notice that for an isotropic velocity distribution, $\beta=0$. In figure \ref{fig:beta} we show the anisotropy parameter as a function of Galactocentric radius for the Auriga MW-like halos. We can see that at small radii, relative velocity distributions of all halos are close to isotropic, but become more anisotropic as we move further from the Galactic center.
 
To better compare halos Au2 and Au22, we can also study the shape of their halos. In section~\ref{sec:properties} we define the range of  sphericities of all the Auriga MW-like halos at four different radii. For Au2 we have $s (2~{\rm kpc})=0.66$, $s (8~{\rm kpc})=0.72$, $s (20~{\rm kpc})=0.71$, and $s (50~{\rm kpc})=0.63$. For Au22 we have $s (2~{\rm kpc})=0.82$, $s (8~{\rm kpc})=0.86$, $s (20~{\rm kpc})=0.88$, and $s (50~{\rm kpc})=0.86$. Deviations from sphericity can be described by the triaxiality parameter,
\begin{equation}
T = \frac{a^2-b^2}{a^2-c^2},
\end{equation}
where $a \geq b \geq c$ are the three axes of the ellipsoid obtained from the inertia tensor. For very oblate systems, $T \approx 0$, whereas for very prolate systems, $T \approx 1$. For Au2 we have $T (2~{\rm kpc})=0.72$, $T (8~{\rm kpc})=0.46$, $T (20~{\rm kpc})=0.17$, and $T (50~{\rm kpc})=0.12$. For Au22 we have $T (2~{\rm kpc})=0.56$, $T (8~{\rm kpc})=0.30$, $T (20~{\rm kpc})=0.31$, and $T (50~{\rm kpc})=0.44$. Hence, Au2 has a larger deviation from sphericity and is more triaxial compared to Au22.

 \clearpage

\bibliographystyle{JHEP}
\bibliography{./refs}

\end{document}